\documentclass[aps,prl,twocolumn,preprintnumbers,amsmath,amssymb,superscriptaddress]{revtex4-2}
\usepackage{graphicx}
\usepackage{epsfig}
\usepackage{dcolumn}
\usepackage{bm}
\usepackage{ulem}
\usepackage{color}
\usepackage{float}
\usepackage[colorlinks=true,linkcolor=blue,citecolor=blue,urlcolor=blue]{hyperref}%
\usepackage{verbatim}
\hypersetup{citecolor = blue}
\usepackage{subfigure}
\usepackage[ruled,vlined]{algorithm2e}

\def\be{\begin{equation}}       \def\ee{\end{equation}}
\def\bea{\begin{eqnarray}}      \def\eea{\end{eqnarray}}
\def\ba{\begin{array}}
\def\ea{\end{array}}
\def\bnum{\begin{enumerate} }
\def\enum{\end{enumerate}}

\def\=>{\Rightarrow}
\def\>{\rightarrow}

\def\eye2{Fathbb{I}}

\usepackage{ listings }

\def\Eq#1{Eq.~(\ref{#1})}
\def\Fig#1{Fig.~\ref{#1}}

\renewcommand{\>}{\rangle}

\usepackage{listings}
\usepackage{color}
\definecolor{lightgray}{gray}{1}

\begin{document}
\graphicspath{{figures/}}

\title{Emergence of charge-$4e$ superconductivity 
from 2D nematic superconductors}
\author{Xuan Zou}
\thanks{These two authors contributed equally to this work.}
\affiliation{Institute for Advanced Study, Tsinghua University, Beijing 100084, China} 

\author{Zhou-Quan Wan}
\thanks{These two authors contributed equally to this work.}
\affiliation{Center for Computational Quantum Physics, Flatiron Institute, New York, NY, 10010, USA}

\author{Hong Yao}
\email{yaohong@tsinghua.edu.cn}
\affiliation{Institute for Advanced Study, Tsinghua University, Beijing 100084, China} 

\date{\today}

\begin{abstract}
Charge-$4e$ superconductivity is an exotic state of matter that may emerge as a vestigial order from a charge-$2e$ superconductor with multicomponent superconducting order parameters. 
Showing its emergence in a lattice phase model from numerically exact large-scale computations has remained rare.   
Here, we propose a kagome lattice model with a nematic superconducting ground state and show that it supports a rich set of vestigial phases at elevated temperatures, including a charge-$4e$ phase and a quasi-long-range nematic phase, through large-scale Monte Carlo simulations. 
Combining theoretical analysis with Monte Carlo simulations, we uncover the nature of these phases and show that the phase transitions are governed by the proliferation of distinct topological defects: the $(\tfrac{1}{2},0)$ half superconducting vortices, the $(\tfrac{1}{2},\tfrac{1}{2})$ vortices, the $(0,1)$ integer nematic vortices, and domain-wall excitations. 
In particular, we demonstrate that domain-wall proliferation is crucial for the quasi-nematic phase and should be carefully accounted for when analyzing phase transitions associated with vestigial charge-$4e$ order.
\end{abstract}

\maketitle
\textit{Introduction.}—Since the Bardeen-Cooper-Schrieffer (BCS) theory \cite{Bardeen1957} was established as the microscopic foundation of conventional superconductivity, understanding unconventional superconductors has remained a central challenge in condensed matter physics. 
In usual superconductors, superconductivity arises from the phase coherence of charge-$2e$ Cooper pairs. 
However, it is also possible to form charge-$4e$ superconductivity  
by condensing electron quartets that consist of four electrons 
\cite{
Kivelson1990,Doucot2002,Aligia2005,Berg2009,Herland2010}.
More broadly, related composite-ordered phases have been discussed in several contexts, including metallic superfluids~\cite{BabaevSudboAshcroft2004,Babaev2004,
Babaev2005PRL,Sudbo2005PRL}, counterflow or paired superfluids
~\cite{Kuklov2003,Kuklov2004,Radzihovsky2004,Radzihovsky2008}, and four-fermion superfluids~\cite{Ropke1998,Wu2005a,Lecheminant2005PRL}.
As a novel state of matter, charge-$4e$ superconductivity and related composite orders have gained increasing attention in recent years
~\cite{Moore2004,Ko2009,Radzihovsky2009,Radzihovsky2011,Agterberg2020,Barci2011,You2012,Armaitis2013,Fradkin2015,Mross2015,Jiang2017,Jian2021,Fernandes2021,Zeng2021,Liu2023,Gnezdilov2021,Li2022,zhou2022,Wu2023,Hecker2023,soldini2023,pan2024frustrated,Egor2025,sun2023,varma2024,Chirolli2024,Hecker2024,Wu2024,ojajarvi2024,volovik2024,verghis2025,maccari2025,zhang2025,huecker2025,gao2025,gao2026,shi2026,Schrade2022,Banszerus2025,shi2026high,wan2026}.  
This exotic state of matter is inherently strongly correlated; even its mean-field Hamiltonian remains an interacting theory~\cite{Jiang2017}, making it an intriguing topic for study. Despite recent experimental progress~\cite{Ge2022}, the identification of charge-$4e$ order in real materials is still scarce, and many of its fundamental properties remain unresolved.

For 2D charge-$2e$ superconductors, phase fluctuations of superconducting (SC) order are described by the XY model, where a direct transition between a charge-$2e$ SC and the normal state occurs at the Berezinskii-Kosterlitz-Thouless (BKT) transition via vortex proliferation \cite{Berezinskii1970,Kosterlitz1973,Kosterlitz1974}. 
In the bosonic phase description, the elementary boson can be interpreted as a Cooper pair carrying charge $2e$, so that condensation of boson pairs corresponds to charge-$4e$ superconductivity. 
In a generalized XY model, paired superfluid order can arise when half vortices play a similar role to normal vortices \cite{Korshunov1985a,Lee1985}.
At sufficiently low temperatures, half vortices are bound by string-like domain-wall defects. 
When the temperature becomes comparable to the string tension, 
domain-wall proliferation drives an Ising transition from elementary to paired superfluid order. With the Cooper pair interpretation introduced above, this corresponds to a transition from charge-$2e$ to charge-$4e$ superconductivity.
At higher temperatures, the proliferation of half vortices disrupts the charge-$4e$ phase. 
This mechanism has been corroborated by field-theoretical analyses, Monte Carlo simulations, and tensor-network calculations \cite{Carpenter1989,Poderoso2011,Shi2011,Hubscher2013,Serna2017,Nui2018a,Song2021,Song2022,Victor2022,Song_2025,Tongyu2025}. A three-dimensional generalization with vortex-loop defects has also been formulated~\cite{Gao2022}.

Alternatively, charge-$4e$ superconductivity has been proposed to arise from SC states with additional broken symmetries, including pair-density-wave (PDW) order \cite{Agterberg2008,Berg2009,Radzihovsky2009,Radzihovsky2011,Agterberg2020,wang2015,Hamidian2016,Rajasekaran2018,Ruan2018,Edkins2019,Du2020,Chen2021} and nematic superconductors \cite{Matano2016,Yonezawa2017,Cao2021,liu2024nematic}.
In these cases, charge-$4e$ order can appear as a vestigial phase \cite{Fradkin2015} when the additional broken symmetry is restored first, with transitions mediated by topological defects such as double dislocations \cite{Agterberg2008,Berg2009,Radzihovsky2009,Radzihovsky2011,Agterberg2020} or nematic vortices \cite{Jian2021,Fernandes2021}.
(Similar mechanisms have also been discussed in the context of related composite orders~\cite{Babaev2004,BabaevSudboAshcroft2004,Bojesen2013,Bojesen2014}).
Several studies have analyzed the phase diagram using renormalization-group analysis and numerical simulations \cite{Jian2021,Fernandes2021,Zeng2021,Liu2023}. 
Nevertheless, the interplay between the domain-wall excitations and point-like topological defects in the emergence of charge-$4e$ order through thermal melting has received comparatively little attention. Since such string-like domain-wall defects can strongly influence the stability of intermediate phases, their contribution requires further theoretical clarification.

\begin{figure}[t]
\includegraphics[width=1.0 \columnwidth]{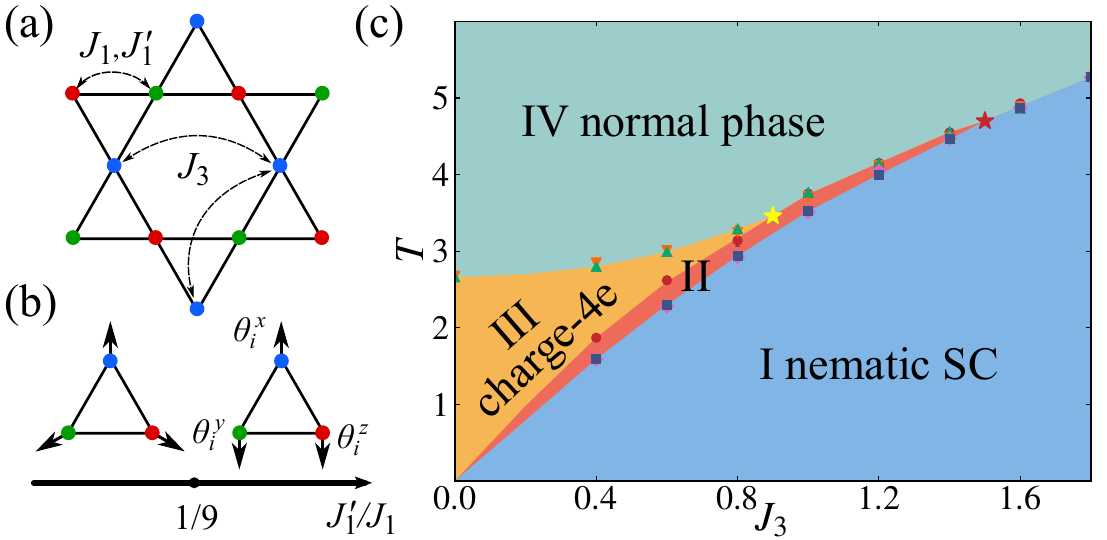}
\caption{(a) Kagome lattice with fields $\theta_i$ residing on the sites and Josephson couplings $J_1$, $J_1'$, and $J_3$.  
(b) Representative ground-state configurations for different ratios of $J_1'/J_1$, up to a global $U(1)$ phase rotation.
(c) Phase diagram of the generalized XY model with $J_1=1$ and $J_1'=3.2$. 
The yellow star at $(J_3^*,T^*)$ and the red star at $(\tilde{J}_3,\tilde{T})$ denote two multicritical points.
Phase I: nematic SC state with charge-$2e$ QLRO and nematic LRO;  
Phase II: nematic charge-$2e$ QLRO with restoration of discrete rotational symmetry, resulting in a quasi-long-range nematic phase;
Phase III: charge-$4e$ QLRO without nematicity;  
Phase IV: disordered phase.  
Colored points denote transition points obtained from different physical observables \cite{SM}.
}
\label{phase}
\end{figure}

In this paper, we propose a generalized XY model whose ground state simultaneously breaks $U(1)$ symmetry and lattice rotational symmetry, realizing a nematic superconductor. Consistent with earlier studies \cite{Jian2021,Fernandes2021}, a charge-$4e$ phase emerges upon heating. However, our Monte Carlo simulations and theoretical analysis reveal a qualitatively \textit{new} phase diagram, featuring an intermediate quasi-long-range nematic phase between the nematic SC and charge-$4e$ states. The appearance of the quasi-nematic order is intriguing, as it contrasts with the conventional expectation that discrete lattice-rotational symmetry breaking in two dimensions (such as reducing $C_4$ or $C_6$ to a lower subgroup, corresponding to an Ising- or $\mathbb{Z}_3$-type order parameter) should produce true long-range nematicity, with exceptions proposed in quasicrystals \cite{liu2024nematic} and twisted bilayer systems \cite{Gali2024}.
In our model, the emergence of the quasi-long-range nematic phase originates from domain-wall excitations, demonstrating that both string-type and vortex-type defects are essential for a complete description of charge-$4e$ melting transitions.

\textit{The model.}—We consider a generalized XY model with $U(1)$ phase fields $\theta_i$ defined on the sites of a kagome lattice, 
as illustrated in \Fig{phase}(a). The Hamiltonian is defined as follows:
\begin{equation}
\begin{split}
H&=J_1\sum_{\langle i,j \rangle}\cos(\theta_i-\theta_j) -J_1'\sum_{\langle i,j \rangle}\cos(2\theta_i-2\theta_j)\\&-J_3\sum_{\langle i,j \rangle'}\cos(\theta_i-\theta_j),
\end{split}\label{gxymodel}
\end{equation}
with $0 \leq \theta_i < 2\pi$. 
Since we focus on zero-field thermal melting, coupling to the electromagnetic vector potential is neglected.
Here, $J_1$ and $J_1'$ denote nearest-neighbor couplings, and $J_3$ denotes the third-nearest-neighbor coupling. Throughout, we assume $J_1, J_1', J_3 > 0$ and omit next-nearest-neighbor couplings. 
Sufficiently weak next-nearest-neighbor couplings may slightly shift the phase boundaries 
but leave the qualitative phase diagram unchanged.
The model respects several symmetries: global $U(1)$ rotations ($\theta_i \to \theta_i + \delta\theta$), time-reversal ($\theta_i \to -\theta_i$), lattice translations, and $C_3$ rotational symmetry.
Josephson-junction arrays~\cite{Schrade2022,Banszerus2025} could provide a promising setting for realizing the model under study in the present paper. 

The ground state of this system can be determined straightforwardly. The kagome lattice consists of three sublattices, indicated by different colors in \Fig{phase}(a). Sites within each sublattice are connected by the $J_3$ couplings without frustration. The ground state is therefore obtained by minimizing the energy within a single triangle, yielding a $120^\circ$ state for $J_1' < J_1/9$ and a nematic state for $J_1' > J_1/9$, as illustrated in \Fig{phase}(b). The configuration shown is a representative ground state since all states related by a global $U(1)$ phase rotation are degenerate.
In this work, we focus on the latter case. This ground state preserves time-reversal symmetry but breaks both $U(1)$ and lattice rotational symmetries and can thus be regarded as a nematic SC state. Specifically, we assume 
the pairing amplitude is already well developed, so that the low-energy fluctuations are predominantly phase fluctuations.
When quantum fluctuations are sufficiently weak, the resulting finite-temperature transitions can be effectively described by the model considered here.

To study the finite-temperature behavior, we performed Monte Carlo simulations to obtain the phase diagram shown in \Fig{phase}(c). For $J_3<\tilde{J}_3$, upon heating, the isotropic SC order, i.e., charge-$2e$ quasi-long-range order (QLRO), is first destroyed at $T_{\text{dw}}$ (phase I $\to$ II), while the nematic long-range order (LRO) simultaneously reduces to QLRO. At higher temperatures, two distinct transition sequences appear, separated by a multicritical point $J_3^*$. For $0<J_3<J_3^*$, nematic QLRO vanishes at $T_{\text{nem}}$ (phase II $\to$ III), followed by the loss of charge-$4e$ QLRO at $T_{4e}$ (phase III $\to$ IV). In contrast, for $J_3^*<J_3<\tilde{J}_3$, the simulations indicate a direct transition from phase II to IV. For larger $J_3>\tilde{J}_3$, the system undergoes a direct transition from phase I to phase IV, in which all orders are destroyed simultaneously.

In the next section, we analyze the topological defects of the model to account for the transition behavior observed in \Fig{phase}(c).

\textit{Effective theory.}—Deep in the nematic superconducting phase, amplitude fluctuations of the order parameters can be neglected. The model described in \Eq{gxymodel} can be characterized by the following order parameters:
\begin{equation}
\begin{aligned}
\Delta_0(\mathbf{r}) &\sim   e^{i\theta_\mathbf{r}^x}+e^{i\theta_\mathbf{r}^y}+e^{i\theta_\mathbf{r}^z},\\
\Delta_{+}(\mathbf{r})& \sim   e^{i\theta_\mathbf{r}^x}+e^{i\left(\frac{2\pi}{3}+\theta_\mathbf{r}^y\right)}+e^{i\left(\frac{4\pi}{3}+\theta_\mathbf{r}^z\right)},\\
\Delta_{-}(\mathbf{r})& \sim  e^{i\theta_\mathbf{r}^x}+e^{i\left(-\frac{2\pi}{3}+\theta_\mathbf{r}^y\right)}+e^{i\left(-\frac{4\pi}{3}+\theta_\mathbf{r}^z\right)},
\end{aligned}
\end{equation}
where $\Delta_0$ denotes the isotropic SC component and $\Delta_\pm$ represent the nematic SC order parameters.
These order parameters transform under $U(1)$ rotations as 
$\Delta_0 \to \Delta_0 e^{i\delta \theta}$ and $\Delta_\pm \to \Delta_\pm e^{i\delta \theta}$, and under $C_3$ lattice rotations as 
$\Delta_0 \to \Delta_0$ and $\Delta_\pm \to \Delta_\pm e^{\pm i 2\pi/3}$.
Neglecting amplitude fluctuations, we parametrize the order parameters as  
$\Delta_0(\mathbf{r}) = |\Delta_0| e^{i\theta_0(\mathbf{r})}$, 
$\Delta_\pm(\mathbf{r}) = |\Delta_{\pm}| e^{i[\theta(\mathbf{r}) \pm \phi(\mathbf{r})]}$.  
From these, we identify the charge-$4e$ order  
$\Delta_{4e}(\mathbf{r}) = \Delta_+ \Delta_- = |\Delta_{4e}| e^{i2\theta(\mathbf{r})}$,  
and the nematic order
$Q(\mathbf{r}) = \Delta_+ \Delta_-^\dagger = |Q| e^{i2\phi(\mathbf{r})}$.
The Ginzburg--Landau description makes the symmetry structure and the associated composite orders explicit, while the lattice phase model enables direct Monte Carlo simulations.

The corresponding effective Hamiltonian density reads
\begin{equation}
\begin{aligned}
H&=\frac{\rho}{2T}(\nabla \theta)^2+\frac{\rho_0}{2T}(\nabla \theta_0)^2
-g\cos(2\theta_0-2\theta) \\
&+\frac{\kappa}{2T}(\nabla \phi)^2
-g_3 \cos(\theta_0-\theta)\cos(3\phi),
\end{aligned}
\end{equation}
where $\rho$ and $\rho_0$ are the superfluid stiffnesses, $g$ couples the isotropic and nematic SC phases, $\kappa$ is the nematic stiffness, and  $g_3$ encodes the lowest-order anisotropy allowed by symmetry. Gauge invariance under $(\theta,\phi) \to (\theta+\pi,\phi+\pi)$ forbids terms such as $\tilde g_3 \cos(3\phi)$. In the absence of $\theta_0$, this Hamiltonian reduces to that of Refs.~\cite{Berg2009,Jian2021}. The $g$ term originates from the coupling 
$\Delta_0^2(\Delta_+\Delta_- )^*+\text{h.c.}$,  
while the $g_3$ term has the form  
$\Delta_0\Delta_+(\Delta_-^*)^2 + \Delta_0\Delta_-(\Delta_+^*)^2 + \text{h.c.}$.
As we will show, the $g$ and $g_3$ terms play a central role in the emergence of quasi-nematicity.

More generally, if the symmetry transformation is $\Delta_\pm \to e^{\pm i 2\pi/q}\Delta_\pm$ with $q\neq3$, the $g_3$ term is forbidden. In that case, the leading coupling between the isotropic and nematic SC components reduces to $\Delta_0^2(\Delta_+\Delta_-)^*+\text{h.c.}\sim \cos(2\theta_0-2\theta)$. This term alone can drive an Ising transition between the charge-$2e$ and charge-$4e$ phases, accompanied by the disappearance of the isotropic SC order. A similar analysis also applies to the coupling between the uniform SC and PDW states, as discussed in the Supplemental Material (SM) \cite{SM}.

To clarify the origin of the phase boundaries in our phase diagram, we first consider the region where the isotropic superconducting component is disordered ($T>T_{\text{dw}}$). 
In this region, the topological defects can be systematically described by switching to the dual boson variables $\tilde{\theta}$ and $\tilde{\phi}$. The corresponding dual Hamiltonian density takes the form \cite{Berg2009,Jian2021}
\begin{equation}
\begin{aligned}
\tilde H= & \frac{T}{2 \rho}(\partial \tilde{\theta})^2+\frac{T}{2 \kappa}(\partial \tilde{\phi})^2-g_{1,0} \cos 2 \pi \tilde{\theta}\\&-g_{0,1} \cos 2 \pi \tilde{\phi} -g_{\frac{1}{2},\frac{1}{2}} \cos \pi \tilde{\theta} \cos \pi \tilde{\phi}.
\end{aligned}\label{ham1}
\end{equation}
where $g_{1,0}, g_{0,1}$, and $ g_{\frac{1}{2},\frac{1}{2}}$ characterize the strength of 
the $(1,0)$, $(0,1)$, and $(\tfrac{1}{2},\tfrac{1}{2})$ vortices, respectively, corresponding to integer SC vortices, integer nematic vortices, and composite half-vortices.
These defects correspond to different windings of the phase fields: SC (nematic) vortices involve $2\pi$ windings of $\theta$ ($\phi$), with energies $\pi \rho \log L$ and $\pi \kappa \log L$, respectively. In addition, because the order parameters $\Delta_\pm$ are invariant under the discrete transformation $(\theta,\phi)\to(\theta+\pi,\phi+\pi)$, a $(\tfrac{1}{2},\tfrac{1}{2})$ vortex is also allowed, carrying $\pi$ windings in both fields and energy $\tfrac{\pi}{4}(\rho+\kappa)\log L$.  

The competition among these vortex species dictates the phase diagram: from the dual perspective, the sequence of transitions is determined by which vortex type proliferates first. When $\kappa/\rho < 1/3$, the $(0,1)$ nematic vortices proliferate and destroy the nematic order, yielding a charge-$4e$ superconductor. Upon further heating, the $(\tfrac{1}{2},0)$ half SC vortices proliferate and drive the loss of charge-$4e$ order \footnote{Here the nematic field $\phi$ is already disordered, so the string tension that confined half SC vortices vanishes. The destruction of charge-$4e$ order is not driven by $(\tfrac{1}{2},\tfrac{1}{2})$ vortices, since half SC vortices have lower energy; this is consistent with our Monte Carlo simulations.}. For $\kappa/\rho$ slightly above $1/3$, proliferation of $(\tfrac{1}{2},\tfrac{1}{2})$ vortices instead produces a direct transition in which nematic and SC orders vanish simultaneously.  

In our model parameters, $J_3$ is proportional to $\kappa$, so increasing $J_3$ raises $\kappa/\rho$. This explains the phase-boundary structure in Fig.~\ref{phase}(c): for $J_3 < J_3^*$, the system undergoes successive transitions (phase~II $\to$ III $\to$ IV), whereas for $J_3 > J_3^*$, nematic and charge-$4e$ orders disappear together in a single transition (phase~I/II $\to$ IV).
These results are fully consistent with the predictions of the dual Hamiltonian in Eq.~\eqref{ham1}.  

However, this reduced model alone cannot explain the emergence of the quasi-nematic phase (phase~II) observed in the full phase diagram. 
In this phase, the isotropic charge-$2e$ SC component is disordered, $\langle e^{i\theta_0} \rangle = 0$, while the nematic charge-$2e$ SC components $\Delta_{\pm}$ and the composite nematic order $Q(\mathbf{r}) = \Delta_+ \Delta_-^\dagger$ exhibit QLRO.
To capture this feature, we need to include the isotropic superconducting field $\theta_0$ and analyze how its coupling to $\theta$ and $\phi$ reshapes the phase structure.

\begin{figure}[t]
\includegraphics[width=\columnwidth]{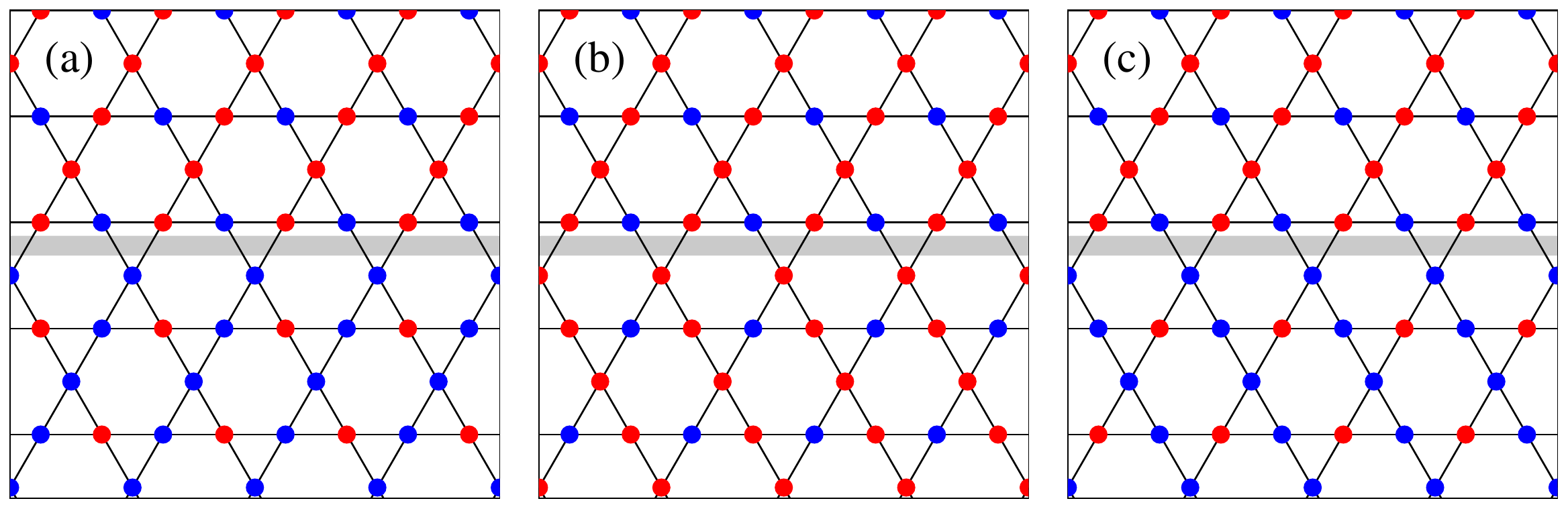}
\caption{Demonstration of different domains with a length of 4 units. The values of $(\theta^x,\theta^y,\theta^z)$ on the three kagome sublattices are $(0,0,\pi)$ in the upper part, and (a) $(\pi,\pi,0)$, (b) $(0,\pi,0)$, and (c) $(\pi,0,\pi)$ below the domain wall (the shaded region). The energy cost of the domain wall increases with its length and is proportional to $6J_3$, $4J_3$, and $2J_3$ for cases (a), (b), and (c), respectively.}
\label{domain}
\end{figure}

\textit{Quasi-nematic phase.}—In general, the isotropic (or uniform) SC component $\Delta_0$ can coexist with multi-component SC states. Experimentally, $\Delta_0$ is often observed together with, and in some cases even dominating over, states such as PDW order \cite{liu2022,zhao_2023,aishwarya_2023}. For $Z_2$-broken nematic superconductivity and incommensurate unidirectional PDW states, the leading coupling between $\Delta_0$ and the multi-component SC fields $\Delta_{\pm}$ takes the form $\Delta_0^2 (\Delta_{+}\Delta_{-})^* + \text{h.c.} \sim \cos(2\theta_0 - 2\theta)$. 
In this case, the transition that disorders the isotropic SC state belongs to the Ising universality class, driven by the proliferation of domain walls.

Related Ising transitions between elementary and paired order have been studied in generalized XY models \cite{Korshunov1985a,Lee1985}.
In contrast, the distinct structure of the present model supports several types of domain walls, as illustrated in \Fig{domain}, and enables the emergence of the new quasi-nematic phase.
Domain wall (a) is of Ising type and induces a $\pi$ shift in the fields $(\theta_0,\theta)\to(\theta_0+\pi,\theta+\pi)$. Domain wall (b) rotates the nematic field by $2\pi/3$, $\phi\to\phi+2\pi/3$. Domain wall (c) combines these effects, inducing $(\theta_0,\theta)\to(\theta_0+\pi,\theta+\pi,\phi\to\phi+2\pi/3)$, or equivalently $(\theta_0\to\theta_0+\pi,\theta\to\theta,\phi\to\phi+\pi/3)$. Among these, type-(c) has the lowest energy cost, and its proliferation drives the system into the quasi-nematic phase.

Specifically, at temperature $T_{\text{dw}}$, the proliferation of type-(c) domain walls disorders the isotropic charge-$2e$ condensate, $\langle e^{i\theta_0}\rangle\simeq 0$, rendering the anisotropic coupling $g_3$ irrelevant. The remaining leading anisotropy is the $\cos(6\phi)$ term, which becomes irrelevant at $T\sim 4T_{\text{nem}}/9$, a scale lower than $T_{\text{dw}}$. As a result, the system directly enters a quasi-long-range nematic phase.
Representative real-space snapshots across $T_{\mathrm{dw}}$, together with larger-size results for the nematic correlation $C(L/2)$, are presented in the SM \cite{SM} and provide direct numerical support for this picture.

In the following section, we present Monte Carlo simulations that provide direct evidence supporting this mechanism and the phase transitions summarized in \Fig{phase}.

\begin{figure}[t]
\includegraphics[trim=0 0 0 0, clip,width=1.0 \columnwidth]{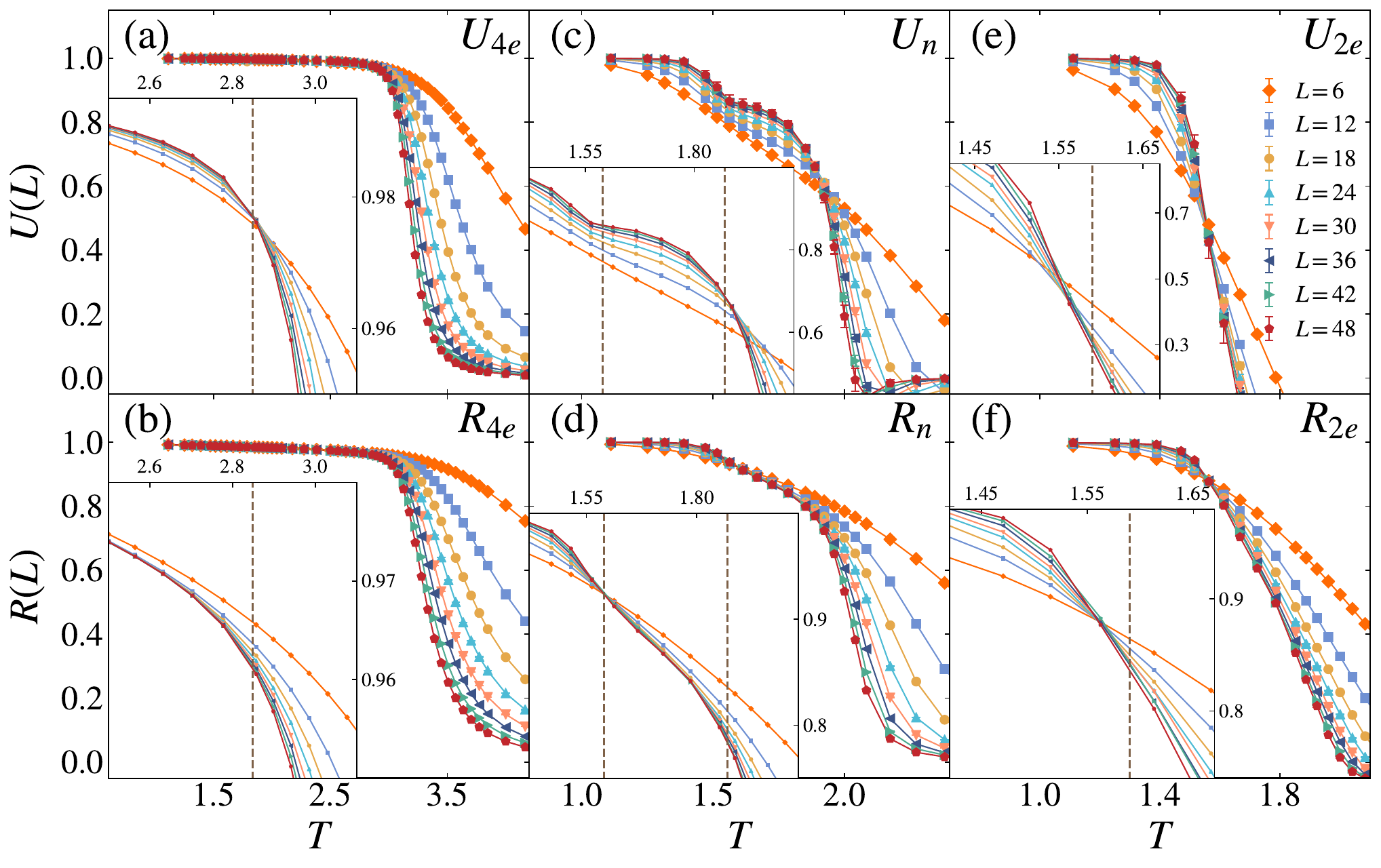}
\caption{The plot of Binder cumulants $U(L)$ and the RG-invariant ratio $R(L)$ for the different orders at $J_3=0.4$. The insets depict a close-up view of the transition points. (a)-(b) correspond to the charge-$4e$ order, where the dashed line denotes the transition temperature $T_{4e}=2.849$.
(c)-(d) represent the nematic order, where the dashed lines denote the transition temperatures $T_{\text{nem}}=1.87$ and $T_{\text{nem}}'\simeq T_{\text{dw}}=1.59$.
(e)-(f) show the charge-$2e$ order, where the dashed line denotes the transition temperature $T_{\text{dw}}=1.59$.
}
\label{J0d4_trans}
\end{figure}

\textit{Monte Carlo results.}—To study the critical behavior and phase diagram, we performed Monte Carlo simulations using the cluster algorithm \cite{Wang1987,Wolff1989}, with $J_1'=3.2$, $J_1=1.0$, and varying $J_3$ to obtain the finite-temperature phase diagram in \Fig{phase}.
The isotropic charge-$2e$ order parameter is  
$\Delta_{2e}(\mathbf{r}_i) = e^{i\theta_i^x} + e^{i\theta_i^y} + e^{i\theta_i^z}$,
with structure factor
$S_{2e}(\mathbf{q},L) = \frac{1}{3L^2}\Big\langle\Big|\sum_{i}\Delta_{2e}(\mathbf{r}_i)\, e^{i\mathbf{q}\cdot \mathbf{r}_i}\Big|^2\Big\rangle$.
Transition points were located using the Binder cumulant and RG-invariant ratio, 
\begin{equation}
    U_{2e} = 2 - \frac{\langle|\sum_i \Delta_{2e}(\mathbf{r}_i)|^4\rangle}
    {\langle|\sum_i \Delta_{2e}(\mathbf{r}_i)|^2\rangle^2}, 
    \quad
    R_{2e} = 1 - \frac{S_{2e}(\delta \mathbf{q},L)}{S_{2e}(0,L)},
\label{binder}
\end{equation}
where $\delta\mathbf{q}$ is the smallest momentum on the finite lattice. The charge-$4e$ and nematic order parameters are defined as  
$\Delta_{4e}(\mathbf{r}_i) = e^{i2\theta_i^x} + e^{i2\theta_i^y} + e^{i2\theta_i^z}, $ and $Q(\mathbf{r}_i) = e^{i(\theta_i^y - \theta_i^z)} 
+ e^{i\frac{2\pi}{3}} e^{i(\theta_i^x - \theta_i^y)} 
+ e^{i\frac{4\pi}{3}} e^{i(\theta_i^z - \theta_i^x)}$, with Binder cumulants and RG ratios defined analogously to Eq.~\eqref{binder}.

In general, the crossing of dimensionless quantities such as Binder cumulants and RG-invariant ratios signals critical points, as these quantities approach unity in the ordered phase and vanish in the disordered phase with increasing system size \cite{Binder1981}. For systems with QLRO, however, the behavior can differ: in a BKT phase, one may still observe crossings, but more characteristically, the curves for different system sizes collapse onto each other, separating only at the transition to the disordered phase.

\Fig{J0d4_trans} shows the behavior of Binder cumulants and the RG-invariant ratios of different order parameters for $J_3=0.4$, with the corresponding results for $J_3=1.2$ and $J_3=1.8$ provided in the SM \cite{SM}.
For charge-$4e$ order, the curves collapse below $T_{4e}$ and separate above it, consistent with QLRO [\Fig{J0d4_trans}(a–b)]. For charge-$2e$ order, a crossing at $T_{\text{dw}}$ signals the loss of isotropic superconductivity [\Fig{J0d4_trans}(e–f)]. For nematic order, Binder cumulants and RG ratios cross at distinct points: they saturate near unity below $T_{\text{nem}}'$, vanish above $T_{\text{nem}}$, and remain finite in the window $T_{\text{nem}}'<T<T_{\text{nem}}$, revealing an intermediate quasi-nematic phase [\Fig{J0d4_trans}(c–d)].

The transition temperatures $T_{4e}$, $T_{\text{dw}}$, $T_{\text{nem}}$, and $T_{\text{nem}}'$ were determined by finite-size extrapolation of the crossing points of $U_{4e}$, $U_{2e}$, $U_{n}$, and $R_{n}$, respectively.
For all the $J_3$ values simulated in this paper, the transition temperature $T_{{\text{nem}}}'$ coincides with $T_{\text{dw}}$ within the numerical errors \cite{SM}, as expected.
Furthermore, for $J_3 > J_3^*$, $T_{4e}$ coincides with $T_{\text{nem}}$, demonstrating the simultaneous destruction of the charge-$4e$ and the nematic QLRO. These results point to a multicritical point $J_3^*$ that separates two distinct transition sequences summarized in \Fig{phase}.

\begin{figure}[t]
\includegraphics[width=0.9 \columnwidth]{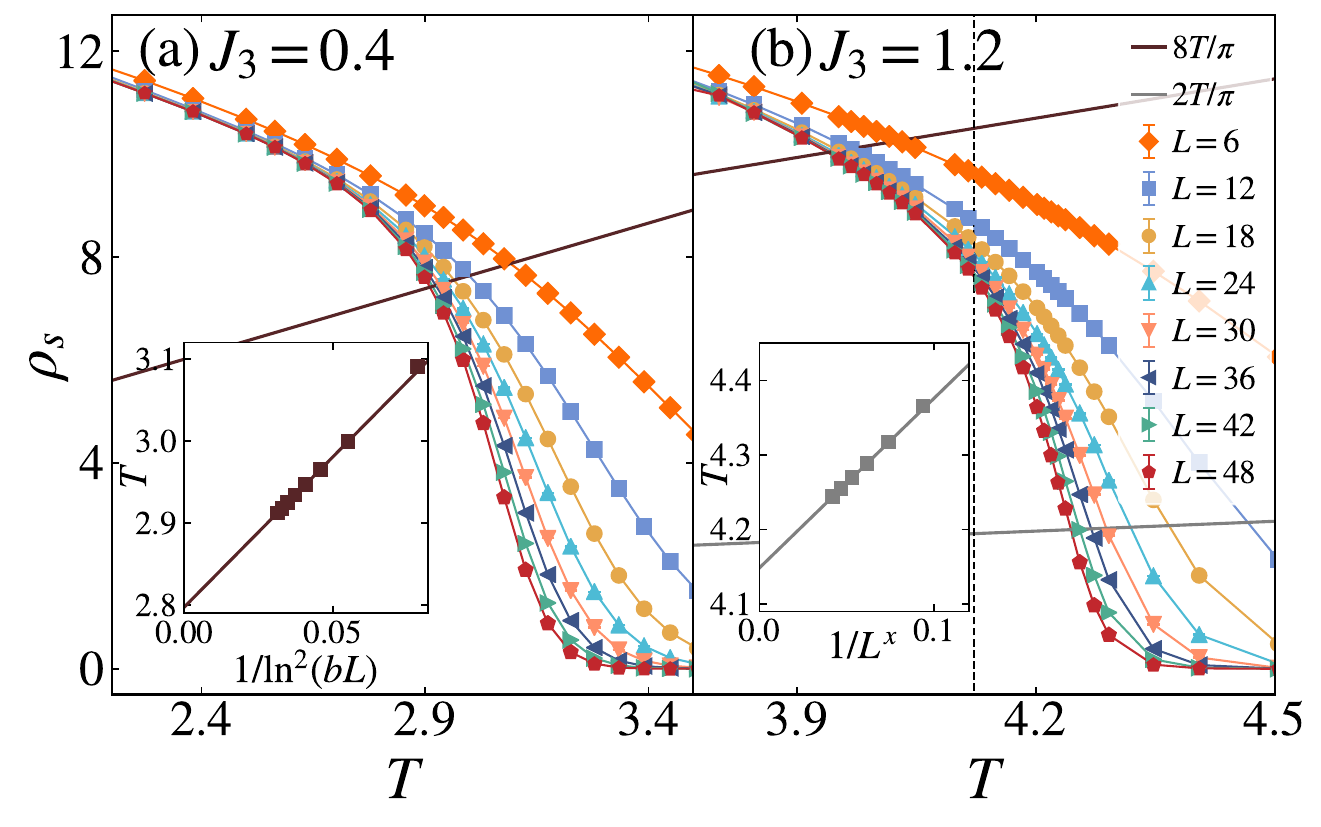}
\caption{(a). The size-dependence of spin stiffness $\rho$ for $J_3=0.4$. The straight line is $\rho=8T/\pi$. The inset figure is the finite-size extrapolation of the critical temperature $T^*(L)$ using $T^*(L)=T_{4e}(\infty)+\frac{a}{\ln^2(bL)}$, where $T^*(L)$ is the crossing point with line $\rho=8 T/\pi$. (b). The size-dependence of spin stiffness $\rho$ for $J_3=1.2$. The two straight lines are $\rho=8 T/\pi $ and $ 2 T /\pi$. The vertical dashed line represents the critical temperature extracted from $U_{4e}$, indicating a stiffness jump lies between $ 8 T/ \pi $ and $2 T/\pi $. The inset figure is the finite-size extrapolation of the critical temperature $T^*(L)$ using $T^*(L)=T_{4e}(\infty)+\frac{a}{L^x}$, where $T^*(L)$ is the crossing point with line $\rho=2  T/\pi$.}
\label{stf}
\end{figure}

To identify which vortices proliferate and to determine the charge-$4e$ transition more precisely, we compute the superfluid stiffness 
$\rho = \partial^2 f(\varphi)/\partial \varphi^2$, 
where $\varphi$ is a twist angle and $f(\varphi)$ is the free-energy density \cite{Hubscher2013}. 
As shown in \Fig{stf}, $\rho$ exhibits the characteristic finite-size scaling behavior of a universal jump at $T_{4e}$. 
For $J_3<J_3^*$, the jump takes the expected value $8T_{4e}/\pi$, consistent with the proliferation of the $(\tfrac{1}{2},0)$ half SC vortices. 
In \Fig{stf}(a), the slope of the straight line is $8/\pi$ and the $T_{4e}$ can be extracted using the finite-size scaling form  
$T^*(L)=T_{4e}(\infty)+a/\ln^2(bL)$ \cite{Weber1988,Hsieh2013}. 
The fit for $J_3=0.4$, shown in the inset of \Fig{stf}(a), yields a transition temperature consistent with that obtained from $U_{4e}$ \cite{SM}. 
For $J_3>J_3^*$, the stiffness jump deviates from $8T_{4e}/\pi$. 
For example, at $J_3=1.2$ the jump falls between $8T_{4e}/\pi$ and $2T_{4e}/\pi$ [\Fig{stf}(b)], consistent with the scenario in which $(\tfrac{1}{2},\tfrac{1}{2})$ vortices proliferate, giving $T_{4e}=\tfrac{\pi}{8}(\rho+\kappa)$.

\begin{figure}[t]
\includegraphics[width=0.9 \columnwidth]{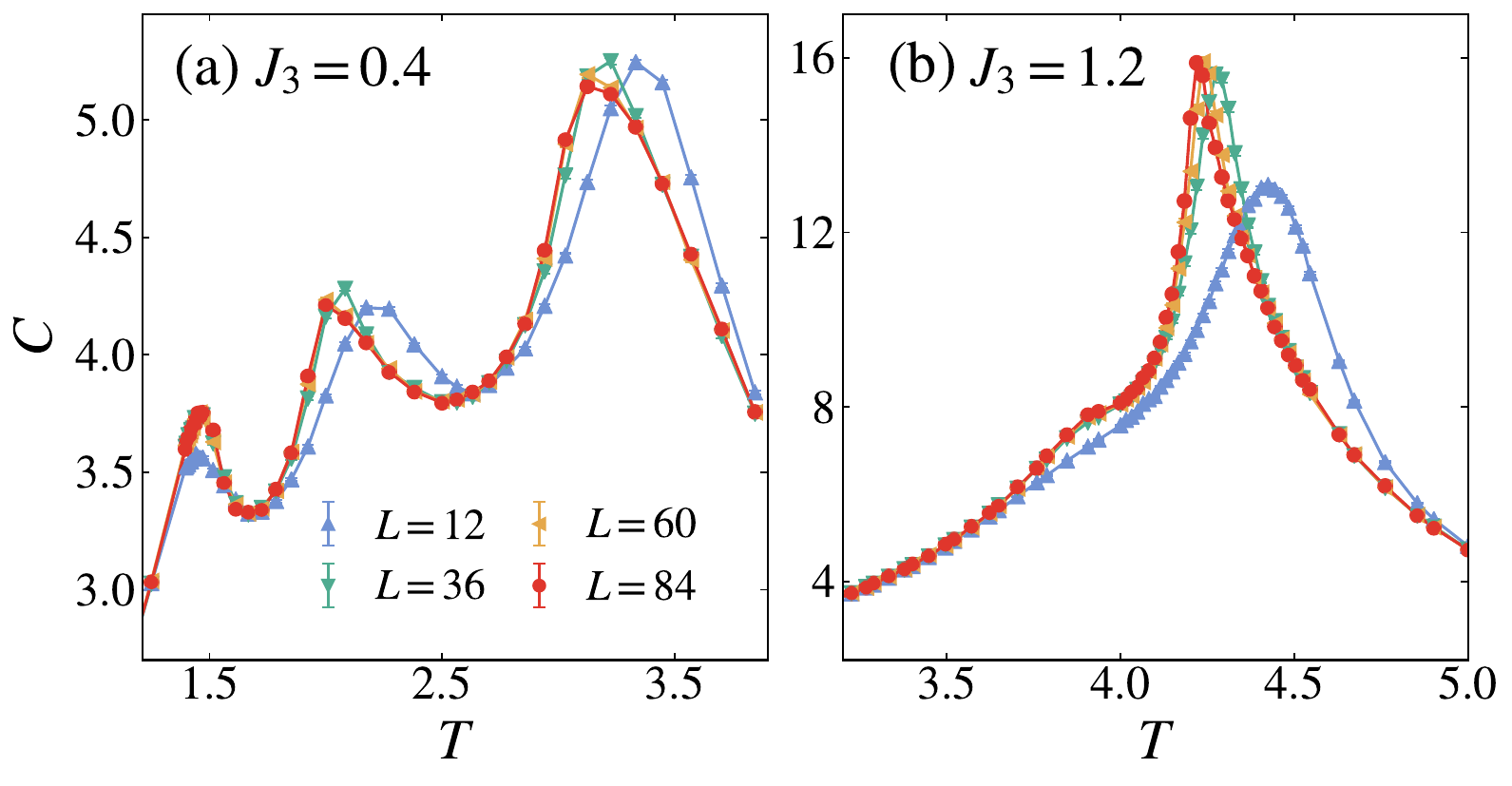}
\caption{The heat capacity for (a) $J_3=0.4$ and (b) $J_3=1.2$. 
The number of bumps indicates the number of transitions, which is three and two, respectively. }
\label{cap}
\end{figure}

To further confirm the two distinct transition behaviors, we computed the heat capacity, shown in \Fig{cap}, which displays three anomalies for $J_3=0.4$ and only two for $J_3=1.2$. 
In conventional generalized XY model, the transition at $T_{\text{dw}}$ is usually attributed to Ising-type domain walls \cite{Lee1985}. 
In our case, the transition at $T_{\text{dw}}$ is not driven by Ising-type domain walls and is therefore unlikely to be Ising. Instead, it is more consistent with a BKT transition: as shown in \Fig{cap}, the heat capacity exhibits a broad bump rather than a sharp peak at $T_{\text{dw}}$.
Within the quasi-long-range nematic phase, the correlation function follows a power law, $\langle \cos(\phi_i-\phi_j)\rangle \sim e^{-\langle(\phi_i-\phi_j)^2\rangle/2} \sim |\mathbf{r}_i-\mathbf{r}_j|^{-\tilde{\eta}(T)}$,  
with $\tilde{\eta}(T_c)=1/4$. Since $Q\sim e^{i2\phi}$, the corresponding critical exponent is $\eta=4\tilde{\eta}$. At $T_{\text{nem}}$, we indeed find $\eta=1$, consistent with the proliferation of the $(0,1)$ nematic vortices \cite{SM}.

\textit{Conclusion.}—To conclude, we have proposed a lattice model that realizes a nematic superconductor and hosts a rich set of vestigial phases, including charge-$4e$ order and quasi-nematicity. The model provides a concrete lattice realization of the field-theoretic scenarios for melting multi-component superconductors into charge-$4e$ states \cite{Berg2009,Radzihovsky2009,Radzihovsky2011,Fernandes2021,Jian2021}. Unlike previous works, we demonstrate that the destruction of charge-$4e$ order is driven by the $(\tfrac{1}{2},0)$ half SC vortices rather than $(\tfrac{1}{2},\tfrac{1}{2})$ vortices, and most strikingly, we identify a new quasi-nematic phase that intervenes between the nematic SC and charge-$4e$ states, stabilized by domain-wall excitations. 

Through theoretical analysis, we show that the sequence of transitions and vestigial orders arises from the proliferation of distinct topological defects, including the $(\tfrac{1}{2},0)$ half SC vortices, the $(\tfrac{1}{2},\tfrac{1}{2})$ vortices, the $(0,1)$ integer nematic vortices, and domain-wall excitations. This picture is further corroborated by Monte Carlo simulations. Our results highlight the crucial role played by domain walls, in addition to various vortices, in shaping the vestigial phases of multi-component superconductors. 
More broadly, it will be interesting to explore how superconducting domain walls and nematic domains influence fractional-vortex proliferation and shape the phase diagrams of systems with vestigial order. Natural future directions include deriving the phase model from microscopic fermionic theories and incorporating quantum dynamics.

{\it Acknowledgments}: We thank Eduardo Fradkin, Shao-Kai Jian, Steve Kivelson, Zijian Wang, and Hao-Xin Wang for helpful discussions. This work is supported in part by MOSTC under Grant No. 2021YFA1400100 (H.Y.), the NSFC under Grant Nos. 12347107 and 12334003 (X.Z. and H.Y.), and the New Cornerstone Science Foundation through the Xplorer Prize
(H.Y.).  The Flatiron Institute is a division of the Simons Foundation.

\bibliography{ref}

@article{Fradkin2015,
  title = {Colloquium: Theory of intertwined orders in high temperature superconductors},
  author = {Fradkin, Eduardo and Kivelson, Steven A. and Tranquada, John M.},
  journal = {Rev. Mod. Phys.},
  volume = {87},
  issue = {2},
  pages = {457--482},
  numpages = {26},
  year = {2015},
  month = {May},
  publisher = {American Physical Society},
  doi = {10.1103/RevModPhys.87.457},
  url = {https://link.aps.org/doi/10.1103/RevModPhys.87.457}
}

@article{Barci2011,
  title = {Role of nematic fluctuations in the thermal melting of pair-density-wave phases in two-dimensional superconductors},
  author = {Barci, Daniel G. and Fradkin, Eduardo},
  journal = {Phys. Rev. B},
  volume = {83},
  issue = {10},
  pages = {100509},
  numpages = {4},
  year = {2011},
  month = {Mar},
  publisher = {American Physical Society},
  doi = {10.1103/PhysRevB.83.100509},
  url = {https://link.aps.org/doi/10.1103/PhysRevB.83.100509}
}

@article{volovik2024,
	title = {Fermionic {Quartet} and {Vestigial} {Gravity}},
	volume = {119},
	issn = {1090-6487},
	url = {https://doi.org/10.1134/S002136402460006X},
	doi = {10.1134/S002136402460006X},
	number = {4},
	journal = {JETP Letters},
	author = {Volovik, G. E.},
	month = feb,
	year = {2024},
	pages = {330--334},
}

@article{verghis2025,
  title = {Vestigial Pairing from Fluctuating Magnetism and Triplet Superconductivity},
  url = {https://arxiv.org/abs/2510.02474},
  journal = {arXiv:2510.02474},
  author = {Verghis, Yanek and Sedov, Denis and Weßling, Jakob and Poduval, Prathyush P. and Scheurer, Mathias S.},
  month = oct,
  year = {2025},
}

@article{Chirolli2024,
  title = {Cooper quartets in interacting hybrid superconducting systems},
  author = {Chirolli, Luca and Braggio, Alessandro and Giazotto, Francesco},
  journal = {Phys. Rev. Res.},
  volume = {6},
  issue = {3},
  pages = {033171},
  numpages = {9},
  year = {2024},
  month = {Aug},
  publisher = {American Physical Society},
  doi = {10.1103/PhysRevResearch.6.033171},
  url = {https://link.aps.org/doi/10.1103/PhysRevResearch.6.033171}
}

@article{varma2024,
  title = {Extended superconducting fluctuation region and $6e$ and $4e$ flux quantization in a kagome compound with a normal state of $3Q$ order},
  author = {Varma, Chandra M. and Wang, Ziqiang},
  journal = {Phys. Rev. B},
  volume = {108},
  issue = {21},
  pages = {214516},
  numpages = {10},
  year = {2023},
  month = {Dec},
  publisher = {American Physical Society},
  doi = {10.1103/PhysRevB.108.214516},
  url = {https://link.aps.org/doi/10.1103/PhysRevB.108.214516}
}

@article{Victor2022,
  title = {Emergent Potts Order in a Coupled Hexatic-Nematic XY model},
  author = {Drouin-Touchette, Victor and Orth, Peter P. and Coleman, Piers and Chandra, Premala and Lubensky, Tom C.},
  journal = {Phys. Rev. X},
  volume = {12},
  issue = {1},
  pages = {011043},
  numpages = {24},
  year = {2022},
  month = {Mar},
  publisher = {American Physical Society},
  doi = {10.1103/PhysRevX.12.011043},
  url = {https://link.aps.org/doi/10.1103/PhysRevX.12.011043}
}

@article{zhou2022,
	title = {Chern {Fermi} pocket, topological pair density wave, and charge-4e and charge-6e superconductivity in kagomé superconductors},
	volume = {13},
	issn = {2041-1723},
	url = {https://doi.org/10.1038/s41467-022-34832-2},
	doi = {10.1038/s41467-022-34832-2},
	number = {1},
	journal = {Nature Communications},
	author = {Zhou, Sen and Wang, Ziqiang},
	month = nov,
	year = {2022},
	pages = {7288},
}

@article{sun2023,
  title = {Emergence of a diverse array of phases in an exactly solvable model},
  author = {Sun, Zhipeng and Lin, Hai-Qing},
  journal = {Phys. Rev. B},
  volume = {109},
  issue = {11},
  pages = {115108},
  numpages = {12},
  year = {2024},
  month = {Mar},
  publisher = {American Physical Society},
  doi = {10.1103/PhysRevB.109.115108},
  url = {https://link.aps.org/doi/10.1103/PhysRevB.109.115108}
}

@article{Hecker2024,
  title = {Local condensation of charge-$4e$ superconductivity at a nematic domain wall},
  author = {Hecker, Matthias and Fernandes, Rafael M.},
  journal = {Phys. Rev. B},
  volume = {109},
  issue = {13},
  pages = {134514},
  numpages = {8},
  year = {2024},
  month = {Apr},
  publisher = {American Physical Society},
  doi = {10.1103/PhysRevB.109.134514},
  url = {https://link.aps.org/doi/10.1103/PhysRevB.109.134514}
}

@article{Wu2024,
  title = {Possible Sliding Regimes in Twisted Bilayer ${\mathrm{WTe}}_{2}$},
  author = {Wu, Yi-Ming and Murthy, Chaitanya and Kivelson, Steven A.},
  journal = {Phys. Rev. Lett.},
  volume = {133},
  issue = {24},
  pages = {246501},
  numpages = {8},
  year = {2024},
  month = {Dec},
  publisher = {American Physical Society},
  doi = {10.1103/PhysRevLett.133.246501},
  url = {https://link.aps.org/doi/10.1103/PhysRevLett.133.246501}
}

@article{ojajarvi2024,
	title = {Pairing at a single {Van} {Hove} point},
	volume = {9},
	issn = {2397-4648},
	url = {https://doi.org/10.1038/s41535-024-00717-4},
	doi = {10.1038/s41535-024-00717-4},
	number = {1},
	journal = {npj Quantum Materials},
	author = {Ojajärvi, Risto and Chubukov, Andrey V. and Lee, Yueh-Chen and Garst, Markus and Schmalian, Jörg},
	month = dec,
	year = {2024},
	pages = {105},
}

@article{maccari2025,
  title = {Revisiting vestigial order in nematic superconductors: Gauge-field mechanisms and model constraints},
  author = {Maccari, I. and Babaev, E. and Carlstr\"om, J.},
  journal = {Phys. Rev. B},
  volume = {113},
  issue = {1},
  pages = {014501},
  numpages = {11},
  year = {2026},
  month = {Jan},
  publisher = {American Physical Society},
  doi = {10.1103/gccc-rfw4},
  url = {https://link.aps.org/doi/10.1103/gccc-rfw4}
}

@article{zhang2025,
  title = {Charge-4$e$ Anyon Superconductor from Doping $\text{SU}(4)_1$ Chiral Spin Liquid},
  url = {https://arxiv.org/abs/2508.12370},
  journal = {arXiv:2508.12370},
  author = {Zhang, Lu and Zhang, Ya-Hui and Song, Xue-Yang},
  month = aug,
  year = {2025},
}

@article{huecker2025,
  title = {Vestigial $d$-wave charge-$4e$ superconductivity from bidirectional pair density waves},
  author = {Huecker, Ethan and Wang, Yuxuan},
  journal = {Phys. Rev. B},
  volume = {113},
  issue = {4},
  pages = {045121},
  numpages = {13},
  year = {2026},
  month = {Jan},
  publisher = {American Physical Society},
  doi = {10.1103/7lbj-tzw3},
  url = {https://link.aps.org/doi/10.1103/7lbj-tzw3}
}

@article{zhao_2023,
	title = {Smectic pair-density-wave order in {EuRbFe4As4}},
	volume = {618},
	issn = {1476-4687},
	url = {https://doi.org/10.1038/s41586-023-06103-7},
	doi = {10.1038/s41586-023-06103-7},
	number = {7967},
	journal = {Nature},
	author = {Zhao, He and Blackwell, Raymond and Thinel, Morgan and Handa, Taketo and Ishida, Shigeyuki and Zhu, Xiaoyang and Iyo, Akira and Eisaki, Hiroshi and Pasupathy, Abhay N. and Fujita, Kazuhiro},
	month = jun,
	year = {2023},
	pages = {940--945},
}

@article{aishwarya_2023,
	title = {Magnetic-field-sensitive charge density waves in the superconductor {UTe2}},
	volume = {618},
	issn = {1476-4687},
	url = {https://doi.org/10.1038/s41586-023-06005-8},
	doi = {10.1038/s41586-023-06005-8},
	number = {7967},
	journal = {Nature},
	author = {Aishwarya, Anuva and May-Mann, Julian and Raghavan, Arjun and Nie, Laimei and Romanelli, Marisa and Ran, Sheng and Saha, Shanta R. and Paglione, Johnpierre and Butch, Nicholas P. and Fradkin, Eduardo and Madhavan, Vidya},
	month = jun,
	year = {2023},
	pages = {928--933},
}

@article{pan2024frustrated,
	title = {Frustrated superconductivity and sextetting order},
	volume = {67},
	issn = {1869-1927},
	url = {https://doi.org/10.1007/s11433-024-2396-y},
	doi = {10.1007/s11433-024-2396-y},
	number = {8},
	journal = {Science China Physics, Mechanics \& Astronomy},
	author = {Pan, Zhiming and Lu, Chen and Yang, Fan and Wu, Congjun},
	month = jul,
	year = {2024},
	pages = {287412},
}

@article{Song_2025,
doi = {10.1088/0256-307X/42/3/037401},
url = {https://doi.org/10.1088/0256-307X/42/3/037401},
year = {2025},
month = {mar},
publisher = {Chinese Physical Society and IOP Publishing Ltd},
volume = {42},
number = {3},
pages = {037401},
author = {Song, Feng-Feng and Zhang, Guang-Ming},
title = {Phase coherence of charge-6e superconductors with a frustrated Kagome XY antiferromagnet},
journal = {Chin. Phys. Lett.}
}

@article{liu2024nematic,
  title = {Nematic Superconductivity and Its Critical Vestigial Phases in the Quasicrystal},
  author = {Liu, Yu-Bo and Zhou, Jing and Yang, Fan},
  journal = {Phys. Rev. Lett.},
  volume = {133},
  issue = {13},
  pages = {136002},
  numpages = {8},
  year = {2024},
  month = {Sep},
  publisher = {American Physical Society},
  doi = {10.1103/PhysRevLett.133.136002},
  url = {https://link.aps.org/doi/10.1103/PhysRevLett.133.136002}
}

@article{Gali2024,
  title = {Critical Nematic Phase with Pseudogaplike Behavior in Twisted Bilayers},
  author = {Gali, Virginia and Hecker, Matthias and Fernandes, Rafael M.},
  journal = {Phys. Rev. Lett.},
  volume = {133},
  issue = {23},
  pages = {236501},
  numpages = {7},
  year = {2024},
  month = {Dec},
  publisher = {American Physical Society},
  doi = {10.1103/PhysRevLett.133.236501},
  url = {https://link.aps.org/doi/10.1103/PhysRevLett.133.236501}
}

@article{Mross2015,
  title = {Spin- and Pair-Density-Wave Glasses},
  author = {Mross, David F. and Senthil, T.},
  journal = {Phys. Rev. X},
  volume = {5},
  issue = {3},
  pages = {031008},
  numpages = {12},
  year = {2015},
  month = {Jul},
  publisher = {American Physical Society},
  doi = {10.1103/PhysRevX.5.031008},
  url = {https://link.aps.org/doi/10.1103/PhysRevX.5.031008}
}

@article{soldini2023,
  title = {Charge-$4e$ superconductivity in a Hubbard model},
  author = {Soldini, Martina O. and Fischer, Mark H. and Neupert, Titus},
  journal = {Phys. Rev. B},
  volume = {109},
  issue = {21},
  pages = {214509},
  numpages = {16},
  year = {2024},
  month = {Jun},
  publisher = {American Physical Society},
  doi = {10.1103/PhysRevB.109.214509},
  url = {https://link.aps.org/doi/10.1103/PhysRevB.109.214509}
}

@article{Liu2023,
  title={Charge-4e superconductivity and chiral metal in 45-twisted bilayer cuprates and related bilayers},
  author={Liu, Yu-Bo and Zhou, Jing and Wu, Congjun and Yang, Fan},
  journal={Nat. Commun.},
  volume={14},
  number={1},
  DOI={10.1038/s41467-023-43782-2},
  pages={7926},
  year={2023},
  publisher={Nature Publishing Group UK London}
}

@article{Wang1987,
  title = {Nonuniversal critical dynamics in Monte Carlo simulations},
  author = {Swendsen, Robert H. and Wang, Jian-Sheng},
  journal = {Phys. Rev. Lett.},
  volume = {58},
  issue = {2},
  pages = {86--88},
  numpages = {0},
  year = {1987},
  month = {Jan},
  publisher = {American Physical Society},
  doi = {10.1103/PhysRevLett.58.86},
  url = {https://link.aps.org/doi/10.1103/PhysRevLett.58.86}
}

@article{Ruan2018,
	title = {Visualization of the periodic modulation of {Cooper} pairing in a cuprate superconductor},
	volume = {14},
	issn = {1745-2481},
	url = {https://doi.org/10.1038/s41567-018-0276-8},
	doi = {10.1038/s41567-018-0276-8},
	number = {12},
	journal = {Nat. Phys.},
	author = {Ruan, Wei and Li, Xintong and Hu, Cheng and Hao, Zhenqi and Li, Haiwei and Cai, Peng and Zhou, Xingjiang and Lee, Dung-Hai and Wang, Yayu},
	month = dec,
	year = {2018},
	pages = {1178--1182},
}

@article{Binder1981,
  title = {Critical Properties from Monte Carlo Coarse Graining and Renormalization},
  author = {Binder, K.},
  journal = {Phys. Rev. Lett.},
  volume = {47},
  issue = {9},
  pages = {693--696},
  numpages = {0},
  year = {1981},
  month = {Aug},
  publisher = {American Physical Society},
  doi = {10.1103/PhysRevLett.47.693},
  url = {https://link.aps.org/doi/10.1103/PhysRevLett.47.693}
}

@article{Matano2016,
	title = {Spin-rotation symmetry breaking in the superconducting state of {CuxBi2Se3}},
	volume = {12},
	issn = {1745-2481},
	url = {https://doi.org/10.1038/nphys3781},
	doi = {10.1038/nphys3781},
	number = {9},
	journal = {Nat. Phys.},
	author = {Matano, K. and Kriener, M. and Segawa, K. and Ando, Y. and Zheng, Guo-qing},
	month = sep,
	year = {2016},
	pages = {852--854},
}

@article{Weber1988,
  title = {Monte Carlo determination of the critical temperature for the two-dimensional XY model},
  author = {Weber, Hans and Minnhagen, Petter},
  journal = {Phys. Rev. B},
  volume = {37},
  issue = {10},
  pages = {5986--5989},
  numpages = {0},
  year = {1988},
  month = {Apr},
  publisher = {American Physical Society},
  doi = {10.1103/PhysRevB.37.5986},
  url = {https://link.aps.org/doi/10.1103/PhysRevB.37.5986}
}

@article{Yonezawa2017,
	title = {Thermodynamic evidence for nematic superconductivity in {CuxBi2Se3}},
	volume = {13},
	issn = {1745-2481},
	url = {https://doi.org/10.1038/nphys3907},
	doi = {10.1038/nphys3907},
	number = {2},
	journal = {Nat. Phys.},
	author = {Yonezawa, Shingo and Tajiri, Kengo and Nakata, Suguru and Nagai, Yuki and Wang, Zhiwei and Segawa, Kouji and Ando, Yoichi and Maeno, Yoshiteru},
	month = feb,
	year = {2017},
	pages = {123--126},
}

@article{Cao2021,
author = {Yuan Cao  and Daniel Rodan-Legrain  and Jeong Min Park  and Noah F. Q. Yuan  and Kenji Watanabe  and Takashi Taniguchi  and Rafael M. Fernandes  and Liang Fu  and Pablo Jarillo-Herrero },
title = {Nematicity and competing orders in superconducting magic-angle graphene},
journal = {Science},
volume = {372},
number = {6539},
pages = {264-271},
year = {2021},
doi = {10.1126/science.abc2836},
URL = {https://www.science.org/doi/abs/10.1126/science.abc2836}
}

@article{Chen2021,
	title = {Roton pair density wave in a strong-coupling kagome superconductor},
	volume = {599},
	issn = {1476-4687},
	url = {https://doi.org/10.1038/s41586-021-03983-5},
	doi = {10.1038/s41586-021-03983-5},
	number = {7884},
	journal = {Nature},
	author = {Chen, Hui and Yang, Haitao and Hu, Bin and Zhao, Zhen and Yuan, Jie and Xing, Yuqing and Qian, Guojian and Huang, Zihao and Li, Geng and Ye, Yuhan and Ma, Sheng and Ni, Shunli and Zhang, Hua and Yin, Qiangwei and Gong, Chunsheng and Tu, Zhijun and Lei, Hechang and Tan, Hengxin and Zhou, Sen and Shen, Chengmin and Dong, Xiaoli and Yan, Binghai and Wang, Ziqiang and Gao, Hong-Jun},
	month = nov,
	year = {2021},
	pages = {222--228},
}

@article{liu2022,
	title = {Pair density wave state in a monolayer high-{Tc} iron-based superconductor},
	volume = {618},
	issn = {1476-4687},
	url = {https://doi.org/10.1038/s41586-023-06072-x},
	doi = {10.1038/s41586-023-06072-x},
	number = {7967},
	journal = {Nature},
	author = {Liu, Yanzhao and Wei, Tianheng and He, Guanyang and Zhang, Yi and Wang, Ziqiang and Wang, Jian},
	month = jun,
	year = {2023},
	pages = {934--939},
}

@article{Wolff1989,
  title = {Collective Monte Carlo Updating for Spin Systems},
  author = {Wolff, Ulli},
  journal = {Phys. Rev. Lett.},
  volume = {62},
  issue = {4},
  pages = {361--364},
  numpages = {0},
  year = {1989},
  month = {Jan},
  publisher = {American Physical Society},
  doi = {10.1103/PhysRevLett.62.361},
  url = {https://link.aps.org/doi/10.1103/PhysRevLett.62.361}
}

@article{Hsieh2013,
doi = {10.1088/1742-5468/2013/09/P09001},
url = {https://doi.org/10.1088/1742-5468/2013/09/P09001},
year = {2013},
month = {sep},
publisher = {IOP Publishing and SISSA},
volume = {2013},
number = {09},
pages = {P09001},
author = {Hsieh, Yun-Da and Kao, Ying-Jer and Sandvik, Anders W},
title = {Finite-size scaling method for the Berezinskii–Kosterlitz–Thouless transition},
journal = {J. Stat. Mech.: Theory Exp.}
}

@article{Edkins2019,
author = {S. D. Edkins  and A. Kostin  and K. Fujita  and A. P. Mackenzie  and H. Eisaki  and S. Uchida  and Subir Sachdev  and Michael J. Lawler  and E.-A. Kim  and J. C. Séamus Davis  and M. H. Hamidian },
title = {Magnetic field–induced pair density wave state in the cuprate vortex halo},
journal = {Science},
volume = {364},
number = {6444},
pages = {976-980},
year = {2019},
doi = {10.1126/science.aat1773},
URL = {https://www.science.org/doi/abs/10.1126/science.aat1773}
}

@article{Du2020,
	title = {Imaging the energy gap modulations of the cuprate pair-density-wave state},
	volume = {580},
	issn = {1476-4687},
	url = {https://doi.org/10.1038/s41586-020-2143-x},
	doi = {10.1038/s41586-020-2143-x},
	number = {7801},
	journal = {Nature},
	author = {Du, Zengyi and Li, Hui and Joo, Sang Hyun and Donoway, Elizabeth P. and Lee, Jinho and Davis, J. C. Séamus and Gu, Genda and Johnson, Peter D. and Fujita, Kazuhiro},
	month = apr,
	year = {2020},
	pages = {65--70},
}

@article{Rajasekaran2018,
author = {S. Rajasekaran  and J. Okamoto  and L. Mathey  and M. Fechner  and V. Thampy  and G. D. Gu  and A. Cavalleri },
title = {Probing optically silent superfluid stripes in cuprates},
journal = {Science},
volume = {359},
number = {6375},
pages = {575-579},
year = {2018},
doi = {10.1126/science.aan3438},
URL = {https://www.science.org/doi/abs/10.1126/science.aan3438}
}

@article{Hamidian2016,
	title = {Detection of a {Cooper}-pair density wave in {Bi2Sr2CaCu2O8}+x},
	volume = {532},
	issn = {1476-4687},
	url = {https://doi.org/10.1038/nature17411},
	doi = {10.1038/nature17411},
	number = {7599},
	journal = {Nature},
	author = {Hamidian, M. H. and Edkins, S. D. and Joo, Sang Hyun and Kostin, A. and Eisaki, H. and Uchida, S. and Lawler, M. J. and Kim, E.-A. and Mackenzie, A. P. and Fujita, K. and Lee, Jinho and Davis, J. C. Séamus},
	month = apr,
	year = {2016},
	pages = {343--347},
}

@article{Agterberg2008,
	title = {Dislocations and vortices in pair-density-wave superconductors},
	volume = {4},
	issn = {1745-2481},
	url = {https://doi.org/10.1038/nphys999},
	doi = {10.1038/nphys999},
	number = {8},
	journal = {Nat. Phys.},
	author = {Agterberg, D. F. and Tsunetsugu, H.},
	month = aug,
	year = {2008},
	pages = {639--642},
}

@article{Kosterlitz1974,
author = {Kosterlitz, J. M.},
doi = {10.1088/0022-3719/7/6/005},
issn = {00223719},
journal = {J. Phys. C Solid State Phys.},
number = {6},
pages = {1046--1060},
title = {{The critical properties of the two-dimensional xy model}},
volume = {7},
year = {1974}
}

@article{Kosterlitz1973,
author = {Kosterlitz, J. M. and Thouless, D. J.},
doi = {10.1088/0022-3719/6/7/010},
issn = {00223719},
journal = {J. Phys. C Solid State Phys.},
month = {apr},
number = {7},
pages = {1181--1203},
title = {{Ordering, metastability and phase transitions in two-dimensional systems}},
url = {https://linkinghub.elsevier.com/retrieve/pii/0021869373900951 https://iopscience.iop.org/article/10.1088/0022-3719/6/7/010},
volume = {6},
year = {1973}
}

@article{Korshunov1985a,
  author       = {Korshunov, S E},
  title        = {Possible splitting of a phase transition in a 2D XY model},
  url          = {https://www.osti.gov/biblio/5670624},
  journal      = {JETP Lett. (Engl. Transl.)},
  issn         = {ISSN JTPLA},
  volume       = {41:5},
  place        = {United States},
  year         = {1985},
  month        = {03}}

@article{Song2021,
  title = {Hybrid Berezinskii-Kosterlitz-Thouless and Ising topological phase transition in the generalized two-dimensional XY model using tensor networks},
  author = {Song, Feng-Feng and Zhang, Guang-Ming},
  journal = {Phys. Rev. B},
  volume = {103},
  issue = {2},
  pages = {024518},
  numpages = {5},
  year = {2021},
  month = {Jan},
  publisher = {American Physical Society},
  doi = {10.1103/PhysRevB.103.024518},
  url = {https://link.aps.org/doi/10.1103/PhysRevB.103.024518}
}

@article{Song2022,
  title = {Phase Coherence of Pairs of Cooper Pairs as Quasi-Long-Range Order of Half-Vortex Pairs in a Two-Dimensional Bilayer System},
  author = {Song, Feng-Feng and Zhang, Guang-Ming},
  journal = {Phys. Rev. Lett.},
  volume = {128},
  issue = {19},
  pages = {195301},
  numpages = {6},
  year = {2022},
  month = {May},
  publisher = {American Physical Society},
  doi = {10.1103/PhysRevLett.128.195301},
  url = {https://link.aps.org/doi/10.1103/PhysRevLett.128.195301}
}

@article{You2012,
  title = {Superfluidity of Bosons in Kagome Lattices with Frustration},
  author = {You, Yi-Zhuang and Chen, Zhu and Sun, Xiao-Qi and Zhai, Hui},
  journal = {Phys. Rev. Lett.},
  volume = {109},
  issue = {26},
  pages = {265302},
  numpages = {5},
  year = {2012},
  month = {Dec},
  publisher = {American Physical Society},
  doi = {10.1103/PhysRevLett.109.265302},
  url = {https://link.aps.org/doi/10.1103/PhysRevLett.109.265302}
}

@article{Gnezdilov2021,
  title = {Solvable model for a charge-$4e$ superconductor},
  author = {Gnezdilov, Nikolay V. and Wang, Yuxuan},
  journal = {Phys. Rev. B},
  volume = {106},
  issue = {9},
  pages = {094508},
  numpages = {22},
  year = {2022},
  month = {Sep},
  publisher = {American Physical Society},
  doi = {10.1103/PhysRevB.106.094508},
  url = {https://link.aps.org/doi/10.1103/PhysRevB.106.094508}
}

@article{Berezinskii1970,
    author = "Berezinskii, V. L.",
    title = "{Destruction of long range order in one-dimensional and two-dimensional systems having a continuous symmetry group. I. Classical systems}",
    journal = "Sov. Phys. JETP",
    volume = "32",
    pages = "493--500",
    year = "1971"
}

@article{Hecker2023,
  title = {Cascade of vestigial orders in two-component superconductors: Nematic, ferromagnetic, $s$-wave charge-$4e$, and $d$-wave charge-$4e$ states},
  author = {Hecker, Matthias and Willa, Roland and Schmalian, J\"org and Fernandes, Rafael M.},
  journal = {Phys. Rev. B},
  volume = {107},
  issue = {22},
  pages = {224503},
  numpages = {26},
  year = {2023},
  month = {Jun},
  publisher = {American Physical Society},
  doi = {10.1103/PhysRevB.107.224503},
  url = {https://link.aps.org/doi/10.1103/PhysRevB.107.224503}
}

@article{Bardeen1957,
  title = {Theory of Superconductivity},
  author = {Bardeen, J. and Cooper, L. N. and Schrieffer, J. R.},
  journal = {Phys. Rev.},
  volume = {108},
  issue = {5},
  pages = {1175--1204},
  numpages = {0},
  year = {1957},
  month = {Dec},
  publisher = {American Physical Society},
  doi = {10.1103/PhysRev.108.1175},
  url = {https://link.aps.org/doi/10.1103/PhysRev.108.1175}
}

@article{Wu2023,
	title = {d-wave charge-4e superconductivity from fluctuating pair density waves},
	volume = {9},
	issn = {2397-4648},
	url = {https://doi.org/10.1038/s41535-024-00674-y},
	doi = {10.1038/s41535-024-00674-y},
	number = {1},
	journal = {npj Quantum Materials},
	author = {Wu, Yi-Ming and Wang, Yuxuan},
	month = sep,
	year = {2024},
	pages = {66},
}

@article{Aligia2005,
  title = {Quartet Formation at $(100)/(110)$ Interfaces of $d$-Wave Superconductors},
  author = {Aligia, A. A. and Kampf, A. P. and Mannhart, J.},
  journal = {Phys. Rev. Lett.},
  volume = {94},
  issue = {24},
  pages = {247004},
  numpages = {4},
  year = {2005},
  month = {Jun},
  publisher = {American Physical Society},
  doi = {10.1103/PhysRevLett.94.247004},
  url = {https://link.aps.org/doi/10.1103/PhysRevLett.94.247004}
}

@article{Ko2009,
  title = {Doped kagome system as exotic superconductor},
  author = {Ko, Wing-Ho and Lee, Patrick A. and Wen, Xiao-Gang},
  journal = {Phys. Rev. B},
  volume = {79},
  issue = {21},
  pages = {214502},
  numpages = {13},
  year = {2009},
  month = {Jun},
  publisher = {American Physical Society},
  doi = {10.1103/PhysRevB.79.214502},
  url = {https://link.aps.org/doi/10.1103/PhysRevB.79.214502}
}

@article{Herland2010,
  title = {Phase transitions in a three dimensional $U(1)\ifmmode\times\else\texttimes\fi{}U(1)$ lattice London superconductor: Metallic superfluid and charge-$4e$ superconducting states},
  author = {Herland, Egil V. and Babaev, Egor and Sudb\o{}, Asle},
  journal = {Phys. Rev. B},
  volume = {82},
  issue = {13},
  pages = {134511},
  numpages = {16},
  year = {2010},
  month = {Oct},
  publisher = {American Physical Society},
  doi = {10.1103/PhysRevB.82.134511},
  url = {https://link.aps.org/doi/10.1103/PhysRevB.82.134511}
}

@article{Li2022,
title = {Charge 4e superconductor: A wavefunction approach},
journal = {Sci. Bull.},
volume = {69},
number = {15},
pages = {2328-2331},
year = {2024},
issn = {2095-9273},
doi = {https://doi.org/10.1016/j.scib.2024.06.002},
url = {https://www.sciencedirect.com/science/article/pii/S2095927324003980},
author = {Pengfei Li and Kun Jiang and Jiangping Hu}
}

@article{Doucot2002,
  title = {Pairing of Cooper Pairs in a Fully Frustrated Josephson-Junction Chain},
  author = {Dou\ifmmode \mbox{\c{c}}\else \c{c}\fi{}ot, Benoit and Vidal, Julien},
  journal = {Phys. Rev. Lett.},
  volume = {88},
  issue = {22},
  pages = {227005},
  numpages = {4},
  year = {2002},
  month = {May},
  publisher = {American Physical Society},
  doi = {10.1103/PhysRevLett.88.227005},
  url = {https://link.aps.org/doi/10.1103/PhysRevLett.88.227005}
}

@article{Moore2004,
  title = {Geometric effects on T-breaking in $p+ip$ and $d+id$ superconducting arrays},
  author = {Moore, J. E. and Lee, D.-H.},
  journal = {Phys. Rev. B},
  volume = {69},
  issue = {10},
  pages = {104511},
  numpages = {8},
  year = {2004},
  month = {Mar},
  publisher = {American Physical Society},
  doi = {10.1103/PhysRevB.69.104511},
  url = {https://link.aps.org/doi/10.1103/PhysRevB.69.104511}
}

@article{Ropke1998,
  title = {Four-Particle Condensate in Strongly Coupled Fermion Systems},
  author = {R\"opke, G. and Schnell, A. and Schuck, P. and Nozi\`eres, P.},
  journal = {Phys. Rev. Lett.},
  volume = {80},
  issue = {15},
  pages = {3177--3180},
  numpages = {0},
  year = {1998},
  month = {Apr},
  publisher = {American Physical Society},
  doi = {10.1103/PhysRevLett.80.3177},
  url = {https://link.aps.org/doi/10.1103/PhysRevLett.80.3177}
}

@article{Armaitis2013,
  title = {Quantum Rotor Model for a Bose-Einstein Condensate of Dipolar Molecules},
  author = {Armaitis, J. and Duine, R. A. and Stoof, H. T. C.},
  journal = {Phys. Rev. Lett.},
  volume = {111},
  issue = {21},
  pages = {215301},
  numpages = {5},
  year = {2013},
  month = {Nov},
  publisher = {American Physical Society},
  doi = {10.1103/PhysRevLett.111.215301},
  url = {https://link.aps.org/doi/10.1103/PhysRevLett.111.215301}
}

@article{Zeng2021,
	title = {High-order time-reversal symmetry breaking normal state},
	volume = {67},
	issn = {1869-1927},
	url = {https://doi.org/10.1007/s11433-023-2287-8},
	doi = {10.1007/s11433-023-2287-8},
	number = {3},
	journal = {Science China Physics, Mechanics \& Astronomy},
	author = {Zeng, Meng and Hu, Lun-Hui and Hu, Hong-Ye and You, Yi-Zhuang and Wu, Congjun},
	month = feb,
	year = {2024},
	pages = {237411},
}

@article{Jian2021,
  title = {Charge-$4e$ Superconductivity from Nematic Superconductors in Two and Three Dimensions},
  author = {Jian, Shao-Kai and Huang, Yingyi and Yao, Hong},
  journal = {Phys. Rev. Lett.},
  volume = {127},
  issue = {22},
  pages = {227001},
  numpages = {6},
  year = {2021},
  month = {Nov},
  publisher = {American Physical Society},
  doi = {10.1103/PhysRevLett.127.227001},
  url = {https://link.aps.org/doi/10.1103/PhysRevLett.127.227001}
}

@article{Agterberg2020,
   author = "Agterberg, Daniel F. and Davis, J.C. Séamus and Edkins, Stephen D. and Fradkin, Eduardo and Van Harlingen, Dale J. and Kivelson, Steven A. and Lee, Patrick A. and Radzihovsky, Leo and Tranquada, John M. and Wang, Yuxuan",
   title = "The Physics of Pair-Density Waves: Cuprate Superconductors and Beyond", 
   journal= "Annu. Rev. Condens. Matter Phys.",
   year = "2020",
   volume = "11",
   number = "Volume 11, 2020",
   pages = "231-270",
   doi = "https://doi.org/10.1146/annurev-conmatphys-031119-050711",
   url = "https://www.annualreviews.org/content/journals/10.1146/annurev-conmatphys-031119-050711",
   publisher = "Annual Reviews",
   issn = "1947-5462",
   type = "Journal Article",
  }

@article{Radzihovsky2009,
  title = {Quantum Liquid Crystals in an Imbalanced Fermi Gas: Fluctuations and Fractional Vortices in Larkin-Ovchinnikov States},
  author = {Radzihovsky, Leo and Vishwanath, Ashvin},
  journal = {Phys. Rev. Lett.},
  volume = {103},
  issue = {1},
  pages = {010404},
  numpages = {4},
  year = {2009},
  month = {Jul},
  publisher = {American Physical Society},
  doi = {10.1103/PhysRevLett.103.010404},
  url = {https://link.aps.org/doi/10.1103/PhysRevLett.103.010404}
}

@article{Radzihovsky2011,
  title = {Fluctuations and phase transitions in Larkin-Ovchinnikov liquid-crystal states of a population-imbalanced resonant Fermi gas},
  author = {Radzihovsky, Leo},
  journal = {Phys. Rev. A},
  volume = {84},
  issue = {2},
  pages = {023611},
  numpages = {47},
  year = {2011},
  month = {Aug},
  publisher = {American Physical Society},
  doi = {10.1103/PhysRevA.84.023611},
  url = {https://link.aps.org/doi/10.1103/PhysRevA.84.023611}
}

@article{Poderoso2011,
  title = {New Ordered Phases in a Class of Generalized $XY$ Models},
  author = {Poderoso, F\'abio C. and Arenzon, Jeferson J. and Levin, Yan},
  journal = {Phys. Rev. Lett.},
  volume = {106},
  issue = {6},
  pages = {067202},
  numpages = {4},
  year = {2011},
  month = {Feb},
  publisher = {American Physical Society},
  doi = {10.1103/PhysRevLett.106.067202},
  url = {https://link.aps.org/doi/10.1103/PhysRevLett.106.067202}
}

@article{Shi2011,
  title = {Boson Pairing and Unusual Criticality in a Generalized $XY$ Model},
  author = {Shi, Yifei and Lamacraft, Austen and Fendley, Paul},
  journal = {Phys. Rev. Lett.},
  volume = {107},
  issue = {24},
  pages = {240601},
  numpages = {5},
  year = {2011},
  month = {Dec},
  publisher = {American Physical Society},
  doi = {10.1103/PhysRevLett.107.240601},
  url = {https://link.aps.org/doi/10.1103/PhysRevLett.107.240601}
}

@article{Serna2017,
doi = {10.1088/1751-8121/aa89a1},
url = {https://doi.org/10.1088/1751-8121/aa89a1},
year = {2017},
month = {sep},
publisher = {IOP Publishing},
volume = {50},
number = {42},
pages = {424003},
author = {Serna, Pablo and Chalker, J T and Fendley, Paul},
title = {Deconfinement transitions in a generalised XY model},
journal = {J. Phys. A: Math. Theor.}
}

@article{Hubscher2013,
  title = {Stiffness jump in the generalized $XY$ model on the square lattice},
  author = {H\"ubscher, David M. and Wessel, Stefan},
  journal = {Phys. Rev. E},
  volume = {87},
  issue = {6},
  pages = {062112},
  numpages = {5},
  year = {2013},
  month = {Jun},
  publisher = {American Physical Society},
  doi = {10.1103/PhysRevE.87.062112},
  url = {https://link.aps.org/doi/10.1103/PhysRevE.87.062112}
}

@article{wang2015,
  title = {Interplay between pair- and charge-density-wave orders in underdoped cuprates},
  author = {Wang, Yuxuan and Agterberg, Daniel F. and Chubukov, Andrey},
  journal = {Phys. Rev. B},
  volume = {91},
  issue = {11},
  pages = {115103},
  numpages = {21},
  year = {2015},
  month = {Mar},
  publisher = {American Physical Society},
  doi = {10.1103/PhysRevB.91.115103},
  url = {https://link.aps.org/doi/10.1103/PhysRevB.91.115103}
}

@article{Ge2022,
  title = {Charge-$4e$ and Charge-$6e$ Flux Quantization and Higher Charge Superconductivity in Kagome Superconductor Ring Devices},
  author = {Ge, Jun and Wang, Pinyuan and Xing, Ying and Yin, Qiangwei and Wang, Anqi and Shen, Jie and Lei, Hechang and Wang, Ziqiang and Wang, Jian},
  journal = {Phys. Rev. X},
  volume = {14},
  issue = {2},
  pages = {021025},
  numpages = {19},
  year = {2024},
  month = {May},
  publisher = {American Physical Society},
  doi = {10.1103/PhysRevX.14.021025},
  url = {https://link.aps.org/doi/10.1103/PhysRevX.14.021025}
}

@article{Lee1985,
  title = {Strings in two-dimensional classical XY models},
  author = {Lee, D. H. and Grinstein, G.},
  journal = {Phys. Rev. Lett.},
  volume = {55},
  issue = {5},
  pages = {541--544},
  numpages = {0},
  year = {1985},
  month = {Jul},
  publisher = {American Physical Society},
  doi = {10.1103/PhysRevLett.55.541},
  url = {https://link.aps.org/doi/10.1103/PhysRevLett.55.541}
}

@article{Gao2022,
  title = {Fractional vortices, $\mathrm{Z}_2$ gauge theory, and the confinement-deconfinement transition},
  author = {Gao, Zhi-Qiang and Huang, Yen-Ta and Lee, Dung-Hai},
  journal = {Phys. Rev. B},
  volume = {106},
  issue = {12},
  pages = {L121105},
  numpages = {5},
  year = {2022},
  month = {Sep},
  publisher = {American Physical Society},
  doi = {10.1103/PhysRevB.106.L121105},
  url = {https://link.aps.org/doi/10.1103/PhysRevB.106.L121105}
}

@article{Kivelson1990,
  title = {Doped antiferromagnets in the weak-hopping limit},
  author = {Kivelson, S. A. and Emery, V. J. and Lin, H. Q.},
  journal = {Phys. Rev. B},
  volume = {42},
  issue = {10},
  pages = {6523--6530},
  numpages = {0},
  year = {1990},
  month = {Oct},
  publisher = {American Physical Society},
  doi = {10.1103/PhysRevB.42.6523},
  url = {https://link.aps.org/doi/10.1103/PhysRevB.42.6523}
}

@article{Jiang2017,
  title = {Charge-$4e$ superconductors: A Majorana quantum Monte Carlo study},
  author = {Jiang, Yi-Fan and Li, Zi-Xiang and Kivelson, Steven A. and Yao, Hong},
  journal = {Phys. Rev. B},
  volume = {95},
  issue = {24},
  pages = {241103},
  numpages = {5},
  year = {2017},
  month = {Jun},
  publisher = {American Physical Society},
  doi = {10.1103/PhysRevB.95.241103},
  url = {https://link.aps.org/doi/10.1103/PhysRevB.95.241103}
}

@article{Berg2009,
author = {Berg, Erez and Fradkin, Eduardo and Kivelson, Steven A.},
doi = {10.1038/nphys1389},
issn = {17452481},
journal = {Nat. Phys.},
month = {nov},
number = {11},
pages = {830--833},
publisher = {Nature Publishing Group},
title = {{Charge-4e superconductivity from pair-density-wave order in certain high-temperature superconductors}},
url = {http://www.nature.com/articles/nphys1389},
volume = {5},
year = {2009}
}

@article{Fernandes2021,
  title = {Charge-$4e$ Superconductivity from Multicomponent Nematic Pairing: Application to Twisted Bilayer Graphene},
  author = {Fernandes, Rafael M. and Fu, Liang},
  journal = {Phys. Rev. Lett.},
  volume = {127},
  issue = {4},
  pages = {047001},
  numpages = {6},
  year = {2021},
  month = {Jul},
  publisher = {American Physical Society},
  doi = {10.1103/PhysRevLett.127.047001},
  url = {https://link.aps.org/doi/10.1103/PhysRevLett.127.047001}
}

@article{Carpenter1989,
doi = {10.1088/0953-8984/1/30/004},
url = {https://doi.org/10.1088/0953-8984/1/30/004},
year = {1989},
month = {jul},
publisher = {},
volume = {1},
number = {30},
pages = {4907},
author = {D B Carpenter and J T Chalker},
title = {The phase diagram of a generalised XY model},
journal = {J. Phys.: Condens. Matter}
}

@article{Wu2005a,
  title = {Competing Orders in One-Dimensional Spin-$3/2$ Fermionic Systems},
  author = {Wu, Congjun},
  journal = {Phys. Rev. Lett.},
  volume = {95},
  issue = {26},
  pages = {266404},
  numpages = {4},
  year = {2005},
  month = {Dec},
  publisher = {American Physical Society},
  doi = {10.1103/PhysRevLett.95.266404},
  url = {https://link.aps.org/doi/10.1103/PhysRevLett.95.266404}
}

@article{Nui2018a,
  title = {Correlation length in a generalized two-dimensional XY model},
  author = {Nui, Duong Xuan and Tuan, Le and Trung Kien, Nguyen Duc and Huy, Pham Thanh and Dang, Hung T. and Viet, Dao Xuan},
  journal = {Phys. Rev. B},
  volume = {98},
  issue = {14},
  pages = {144421},
  numpages = {9},
  year = {2018},
  month = {Oct},
  publisher = {American Physical Society},
  doi = {10.1103/PhysRevB.98.144421},
  url = {https://link.aps.org/doi/10.1103/PhysRevB.98.144421}
}

@article{Egor2025,
  title = {Microscopic theory of electron quadrupling condensates},
  author = {Samoilenka, Albert and Babaev, Egor},
  journal = {Phys. Rev. Res.},
  volume = {8},
  issue = {1},
  pages = {013139},
  numpages = {12},
  year = {2026},
  month = {Feb},
  publisher = {American Physical Society},
  doi = {10.1103/bm6w-z2lf},
  url = {https://link.aps.org/doi/10.1103/bm6w-z2lf}
}

@misc{SM,
  note = {
  See the Supplemental Material for the Landau free-energy analysis, additional Monte Carlo data and fitting details, real-space snapshots of domain-wall configurations, and correlation-function analyses.}
}

@article{Radzihovsky2004,
  title = {Superfluid Transitions in Bosonic Atom-Molecule Mixtures near a Feshbach Resonance},
  author = {Radzihovsky, Leo and Park, Jae and Weichman, Peter B.},
  journal = {Phys. Rev. Lett.},
  volume = {92},
  issue = {16},
  pages = {160402},
  numpages = {4},
  year = {2004},
  month = {Apr},
  publisher = {American Physical Society},
  doi = {10.1103/PhysRevLett.92.160402},
  url = {https://link.aps.org/doi/10.1103/PhysRevLett.92.160402}
}

@article{Radzihovsky2008,
title = {Superfluidity and phase transitions in a resonant Bose gas},
journal = {Annals of Physics},
volume = {323},
number = {10},
pages = {2376-2451},
year = {2008},
issn = {0003-4916},
doi = {https://doi.org/10.1016/j.aop.2008.05.008},
url = {https://www.sciencedirect.com/science/article/pii/S0003491608000742},
author = {Leo Radzihovsky and Peter B. Weichman and Jae I. Park}
}

@article{gao2025,
  title = {Topological Charge-2ne Superconductors},
  url = {https://arxiv.org/abs/2512.21325},
  journal = {arXiv:2512.21325},
  author = {Gao, Zhi-Qiang and Wang, Yan-Qi and Yang, Hui and Wu, Congjun},
  year = {2025},
}

@article{gao2026,
  title = {Primary charge-4e superconductivity from doping a featureless Mott insulator},
  url = {https://arxiv.org/abs/2602.03925},
  journal = {arXiv:2602.03925},
  author = {Gao, Zhi-Qiang and Wang, Yan-Qi and Zhang, Ya-Hui and Yang, Hui},
  month = feb,
  year = {2026},
}

@article{shi2026,
  title = {Charge-$4e$ superconductor with parafermionic vortices: A path to universal topological quantum computation},
  url = {https://arxiv.org/abs/2602.06963},
  journal = {arXiv:2602.06963},
  author = {Shi, Zhengyan Darius and Han, Zhaoyu and Raghu, Srinivas and Vishwanath, Ashvin},
  month = feb,
  year = {2026},
}

@article{Schrade2022,
  title = {Protected Hybrid Superconducting Qubit in an Array of Gate-Tunable Josephson Interferometers},
  author = {Schrade, Constantin and Marcus, Charles M. and Gyenis, Andr\'as},
  journal = {PRX Quantum},
  volume = {3},
  issue = {3},
  pages = {030303},
  numpages = {15},
  year = {2022},
  month = {Jul},
  publisher = {American Physical Society},
  doi = {10.1103/PRXQuantum.3.030303},
  url = {https://link.aps.org/doi/10.1103/PRXQuantum.3.030303}
}

@article{Banszerus2025,
  title = {Hybrid Josephson Rhombus: A Superconducting Element with Tailored Current-Phase Relation},
  author = {Banszerus, L. and Andersson, C. W. and Marshall, W. and Lindemann, T. and Manfra, M. J. and Marcus, C. M. and Vaitiek\ifmmode \dot{e}\else \.{e}\fi{}nas, S.},
  journal = {Phys. Rev. X},
  volume = {15},
  issue = {1},
  pages = {011021},
  numpages = {12},
  year = {2025},
  month = {Feb},
  publisher = {American Physical Society},
  doi = {10.1103/PhysRevX.15.011021},
  url = {https://link.aps.org/doi/10.1103/PhysRevX.15.011021}
}

@article{wan2026,
  title = {Quantum Charge-$4e$ Superconductivity and Deconfined Pseudocriticality in the Attractive SU(4) Hubbard Model},
  url = {https://arxiv.org/abs/2604.14289},
  journal = {arXiv:2604.14289},
  author = {Wan, Zhou-Quan and Jiang, Huan and Zou, Xuan and Zhang, Shiwei and Jian, Shao-Kai},
  month = apr,
  year = {2026},
}

@article{shi2026high,
  title = {High-temperature charge-$4e$ superconductivity in SU(4) interacting fermions},
  url = {https://arxiv.org/abs/2604.15056},
  journal = {arXiv:2604.15056},
  author = {Shi, Shao-Hang and Wu, Zhengzhi and Hu, Jiangping and Li, Zi-Xiang},
  month = apr,
  year = {2026},
}

@article{Babaev2004,
title = {Phase diagram of planar U(1)×U(1) superconductor: Condensation of vortices with fractional flux and a superfluid state},
journal = {Nuclear Physics B},
volume = {686},
number = {3},
pages = {397-412},
year = {2004},
issn = {0550-3213},
doi = {https://doi.org/10.1016/j.nuclphysb.2004.02.021},
url = {https://www.sciencedirect.com/science/article/pii/S0550321304001294},
author = {Egor Babaev}
}

@article{Kuklov2003,
  title = {Counterflow Superfluidity of Two-Species Ultracold Atoms in a Commensurate Optical Lattice},
  author = {Kuklov, A. B. and Svistunov, B. V.},
  journal = {Phys. Rev. Lett.},
  volume = {90},
  issue = {10},
  pages = {100401},
  numpages = {4},
  year = {2003},
  month = {Mar},
  publisher = {American Physical Society},
  doi = {10.1103/PhysRevLett.90.100401},
  url = {https://link.aps.org/doi/10.1103/PhysRevLett.90.100401}
}

@article{Kuklov2004,
  title = {Commensurate Two-Component Bosons in an Optical Lattice: Ground State Phase Diagram},
  author = {Kuklov, Anatoly and Prokof'ev, Nikolay and Svistunov, Boris},
  journal = {Phys. Rev. Lett.},
  volume = {92},
  issue = {5},
  pages = {050402},
  numpages = {4},
  year = {2004},
  month = {Feb},
  publisher = {American Physical Society},
  doi = {10.1103/PhysRevLett.92.050402},
  url = {https://link.aps.org/doi/10.1103/PhysRevLett.92.050402}
}

@article{BabaevSudboAshcroft2004,
  title   = {A superconductor-to-superfluid phase transition in liquid metallic hydrogen},
  author  = {Babaev, Egor and Sudb{\o}, Asle and Ashcroft, N. W.},
  journal = {Nature},
  volume  = {431},
  number  = {7009},
  pages   = {666--668},
  year    = {2004},
  doi     = {10.1038/nature02910}
}

@article{Bojesen2013,
  title = {Time reversal symmetry breakdown in normal and superconducting states in frustrated three-band systems},
  author = {Bojesen, Troels Arnfred and Babaev, Egor and Sudb\o{}, Asle},
  journal = {Phys. Rev. B},
  volume = {88},
  issue = {22},
  pages = {220511(R)},
  numpages = {4},
  year = {2013},
  month = {Dec},
  publisher = {American Physical Society},
  doi = {10.1103/PhysRevB.88.220511},
  url = {https://link.aps.org/doi/10.1103/PhysRevB.88.220511}
}

@article{Bojesen2014,
  title = {Phase transitions and anomalous normal state in superconductors with broken time-reversal symmetry},
  author = {Bojesen, Troels Arnfred and Babaev, Egor and Sudb\o{}, Asle},
  journal = {Phys. Rev. B},
  volume = {89},
  issue = {10},
  pages = {104509},
  numpages = {11},
  year = {2014},
  month = {Mar},
  publisher = {American Physical Society},
  doi = {10.1103/PhysRevB.89.104509},
  url = {https://link.aps.org/doi/10.1103/PhysRevB.89.104509}
}

@article{Babaev2005PRL,
  title = {Observability of a Projected New State of Matter: A Metallic Superfluid},
  author = {Babaev, E. and Sudb\o{}, A. and Ashcroft, N. W.},
  journal = {Phys. Rev. Lett.},
  volume = {95},
  issue = {10},
  pages = {105301},
  numpages = {4},
  year = {2005},
  month = {Sep},
  publisher = {American Physical Society},
  doi = {10.1103/PhysRevLett.95.105301},
  url = {https://link.aps.org/doi/10.1103/PhysRevLett.95.105301}
}

@article{Sudbo2005PRL,
  title = {Observation of a Metallic Superfluid in a Numerical Experiment},
  author = {Sm\o{}rgrav, E. and Babaev, E. and Smiseth, J. and Sudb\o{}, A.},
  journal = {Phys. Rev. Lett.},
  volume = {95},
  issue = {13},
  pages = {135301},
  numpages = {4},
  year = {2005},
  month = {Sep},
  publisher = {American Physical Society},
  doi = {10.1103/PhysRevLett.95.135301},
  url = {https://link.aps.org/doi/10.1103/PhysRevLett.95.135301}
}

@article{Lecheminant2005PRL,
  title = {Confinement and Superfluidity in One-Dimensional Degenerate Fermionic Cold Atoms},
  author = {Lecheminant, P. and Boulat, E. and Azaria, P.},
  journal = {Phys. Rev. Lett.},
  volume = {95},
  issue = {24},
  pages = {240402},
  numpages = {4},
  year = {2005},
  month = {Dec},
  publisher = {American Physical Society},
  doi = {10.1103/PhysRevLett.95.240402},
  url = {https://link.aps.org/doi/10.1103/PhysRevLett.95.240402}
}

@article{Tongyu2025,
  title = {Theory of the charge-$6e$ condensed phase in kagome-lattice superconductors},
  author = {Lin, Tong-Yu and Song, Feng-Feng and Zhang, Guang-Ming},
  journal = {Phys. Rev. B},
  volume = {111},
  issue = {5},
  pages = {054508},
  numpages = {16},
  year = {2025},
  month = {Feb},
  publisher = {American Physical Society},
  doi = {10.1103/PhysRevB.111.054508},
  url = {https://link.aps.org/doi/10.1103/PhysRevB.111.054508}
}

\end{document}


\begin{center}
\textbf{\large Supplemental Material for ``Emergence of charge-$4e$ superconductivity 
from 2D nematic superconductors
''}
\end{center}

\author{Xuan Zou}
\thanks{These two authors contributed equally to this work.}
\affiliation{Institute for Advanced Study, Tsinghua University, Beijing 100084, China} 

\author{Zhou-Quan Wan}
\thanks{These two authors contributed equally to this work.}
\affiliation{Center for Computational Quantum Physics, Flatiron Institute, New York, NY, 10010, USA}

\author{Hong Yao}
\email{yaohong@tsinghua.edu.cn}
\affiliation{Institute for Advanced Study, Tsinghua University, Beijing 100084, China} 

\date{\today}

\maketitle

\renewcommand{\thefigure}{S\arabic{figure}}
\setcounter{figure}{0}
\renewcommand{\thetable}{S\arabic{table}}
\setcounter{table}{0}
\renewcommand{\theequation}{S\arabic{equation}}
\setcounter{equation}{0}
\renewcommand{\thesection}{\Roman{section}}
\setcounter{section}{0}
\setcounter{secnumdepth}{4}

In this Supplemental Material, we present the Landau free energy analysis, additional Monte Carlo data and fitting details, real-space snapshots of domain-wall configurations, and correlation-function analyses, which further support the conclusions of the main text.
\subsection{Landau free energy}
The coupling term involving the fields $\theta_0$, $\theta$, and $\phi$ in the free energy is given by:
\begin{eqnarray*}
 \mathcal{F}&=& \frac{g}{2}\Delta_0^2 \Delta_+^* \Delta_-^*+\frac{g_3}{4}\Delta_0 \Delta_+ (\Delta_-^*)^2+\frac{g_3}{4}\Delta_0 \Delta_- (\Delta_+^*)^2+\frac{g_6}{2}(\Delta_{+}\Delta_{-}^*)^3+\text{h.c.}
 \\
&=& g|\Delta_0|^2 |\Delta_\pm|^2 \cos(2\theta_0-2\theta)+\frac{g_3}{2} |\Delta_0| |\Delta_\pm|^3\cos(\theta_0-\theta+3\phi)\\
&&+\frac{g_3}{2}|\Delta_0| |\Delta_\pm|^3\cos(\theta_0-\theta-3\phi)+g_6 |\Delta_\pm|^6\cos(6\phi)
\\
&=& g|\Delta_0|^2 |\Delta_\pm|^2 \cos(2\theta_0-2\theta)+g_3|\Delta_0| |\Delta_\pm|^3\cos(\theta_0-\theta)\cos(3\phi)+g_6 |\Delta_\pm|^6\cos(6\phi)
\end{eqnarray*}

Note that the representation $\Delta_\pm = |\Delta_{\pm}| e^{i(\theta \pm \phi)}$ exhibits a $\mathbb{Z}_2$ gauge redundancy under the transformations $\theta \to \theta + \pi$ and $\phi \to \phi + \pi$. Consequently, a term like $g_3 \cos(3\phi)$ cannot appear independently in the free energy; it must be coupled with a $\theta$-dependent term to ensure gauge invariance.
As discussed in the main text, the $\cos(6\phi)$ term becomes irrelevant at a temperature $T \sim \frac{4T_{\text{nem}}}{9}$, which is lower than $T_{\text{nem}}'\simeq T_{\text{dw}}$.

The transition from Phase I to Phase II corresponds to the disordering of the $\Delta_0$ field. In other words, as the temperature increases past $T_{\text{dw}}$, the $\mathbb{Z}_2$ symmetry of the $\Delta_0$ field is restored, and the discrete symmetry of the $\phi$ field is enlarged to a $U(1)$ symmetry. Consequently, the isotropic charge-$2e$ component vanishes, and the nematic order transitions into a quasi-long-range (QLR) order.

The coupling between $\Delta_0$ and $\Delta_\pm$ can also be viewed in the context of PDW superconductivity. In $\text{La}_{2-x}\text{Ba}_x\text{CuO}_4$ near $x=1/8$, one possible viewpoint is that the low-temperature superconducting state may involve both PDW and uniform components. As the temperature increases and the uniform component is suppressed, the system may then evolve toward a more PDW-dominated state, which is related to disordering the $\Delta_0$ field. In the PDW case, however, no lattice-anisotropic coupling to $\Delta_0$ exists (i.e., no $g_3$ term), and the corresponding transition is expected to follow a $\mathbb{Z}_2$ character rather than the behavior discussed here.


\subsection{Finite-size fitting details}
For $J_3\le0.8$, there is a universal jump in the superfluid stiffness $\rho$ at the BKT transition induced by half vortices proliferation, and the stiffness is related to the transition temperature $T_{4e}$ by the equation $\rho=\frac{8}{\pi}T_{4e}$. To determine the transition temperature at the thermodynamic limit, the fitting function $T^*(L)=T_{4e}(\infty)+\frac{a}{\ln^2(bL)}$ is applied, where $L$ is the system size.

For $J_3 \geq 1.0$, the superfluid stiffness no longer exhibits the universal jump at $\rho = \tfrac{8}{\pi}T_c$, but instead falls between $\rho = \tfrac{8}{\pi}T_c$ and $\rho = \tfrac{2}{\pi}T_c$. 
In this regime, the relation $\rho = \tfrac{2}{\pi}T_{4e}$ is used for fitting, which lies below the jump and can be extrapolated to obtain the correct transition temperature $T_{4e}$. In this case, the transition temperature is determined using the general function $T^*(L)=T_{4e}(\infty)+b/L^x$.

As for the dimensionless quantities Binder Cumulants and the RG invariant ratio, the crossing points of the data with size $L$ and $L+6$ are first obtained. Then, the finite-size extrapolation is applied with the power-law form $T_c(L)=T_c(\infty)+b/L^x$.
All the fitting results are summarized in Tab. \ref{tab1}. 

\begin{table}[h]
  \centering
  \caption{The transition temperatures with different values of $J_3$ are determined from different physical quantities. The stiffness $\rho$ is the superfluid stiffness, which exhibits a discontinuity at $T_{4e}$. The $U_{4e}$, $U_{2e}$, and $U_{n}$ are the Binder cumulants of charge-$4e$, isotropic charge-$2e$, and nematic orders. $R_{n}$ is the RG-invariant ratio of nematic order. $T_{\text{dw}}$, $T_{\text{nem}}$, and $T_{4e}$ are related to the transition temperatures from phase I to II, II to III, and III to IV, respectively. $T_{\text{nem}}'$ is the transition from nematic long-range order to quasi-long-range order, which coincides with $T_{\text{dw}}.$}
    \begin{tabular}{ccccccccccc}
    \hline
    \hline
    $T$ & $J_3$                   & 0                & 0.4              & 0.6            & 0.8            & 1             &1.2          & 1.4  & 1.6           &1.8        \\
    \hline
$T_{4e}$ &     $\rho$      & 2.6597(7)  & 2.7971(11) & 2.9915(6) & 3.2865(2) & 3.763(3) & 4.149(8) & 4.555(9) &4.86(3) & 5.277(8) \\
$T_{4e}$ &     $U_{4e}$ & 2.678(8)    & 2.849(3)     & 3.01(4)      & 3.25(8)     & 3.72(6)   &4.123(5)  & 4.493(5) &4.891(4) &5.270(5)\\
$T_{\text{dw}}$ &     $U_{2e}$ &   --              & 1.59(4)       & 2.28(3)      & 2.93(4)     & 3.51(4)   &4.05(3) & 4.47(1) &4.874(3)   &5.268(5)\\
$T_{\text{nem}}$ &    $U_n$     &   --              & 1.87(2)       & 2.620(6)    & 3.14(8)     & 3.75(1)   &4.15(4)    & 4.55(5) &4.93(4)  &5.28(3)  \\
$T_{\text{nem}}'$ &     $R_n$     &   --              & 1.59(1)       & 2.30(1)      & 2.936(4)   & 3.52(2)   &4.000(3)  & 4.459(8) &4.872(2) &5.269(2)\\
    \hline
    \hline
    \end{tabular}%
  \label{tab1}%
\end{table}%

\begin{figure}[t]
\includegraphics[trim=0 0 0 0, clip,width=0.8 \columnwidth]{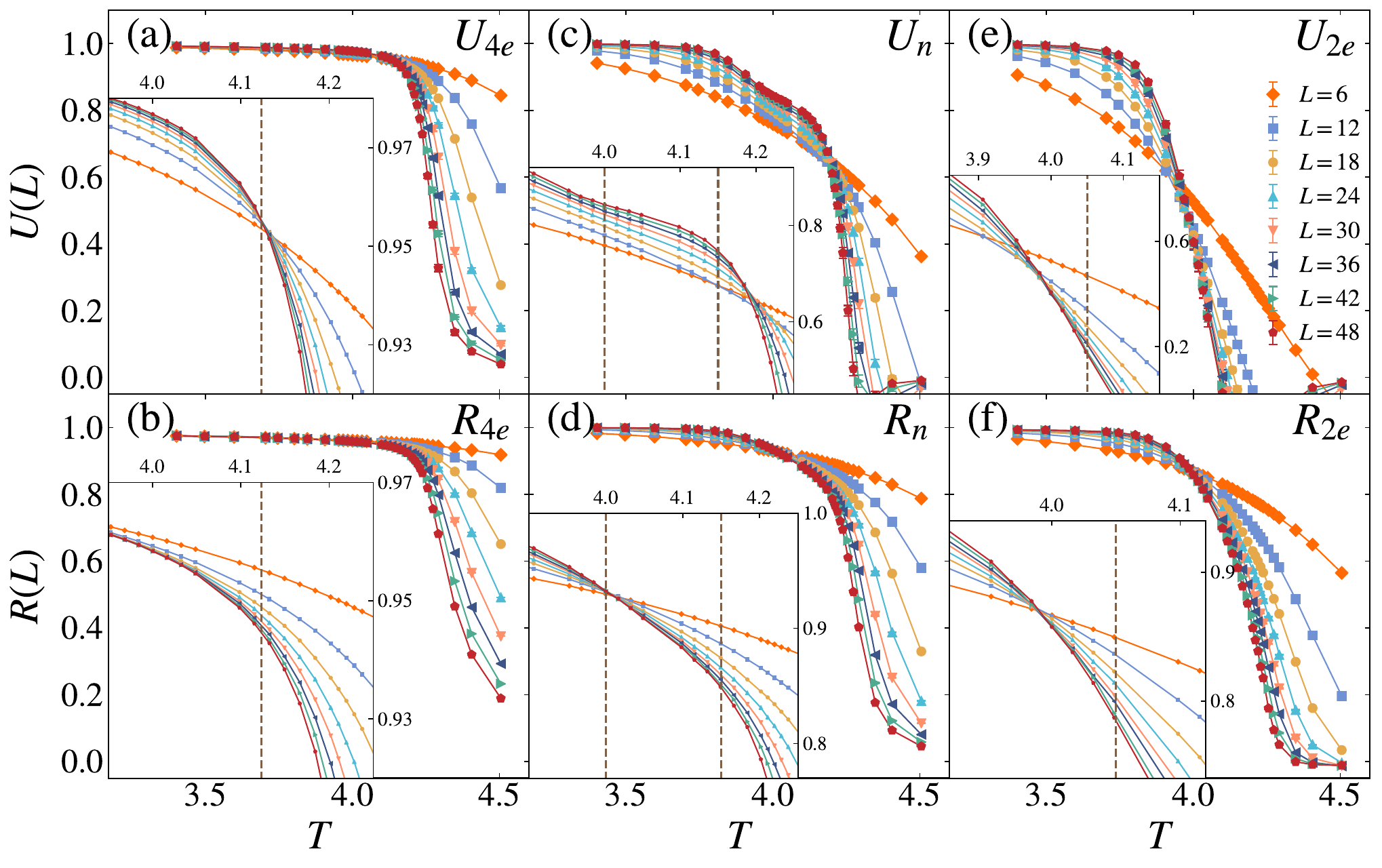}
\caption{
The plot of Binder cumulants $U(L)$ and the RG-invariant ratio $R(L)$ for the different orders at $J_3=1.2$. The insets depict a close-up view of the transition points. (a)-(b) correspond to the charge-$4e$ order, where the dashed line denotes the transition temperature $T_{4e}=4.123$.
(c)-(d) represent the nematic order, with the dashed lines indicating the transition temperatures $T_{\text{nem}}=4.15$ and $T_{\text{nem}}'=4.0$.
(e)-(f) show the isotropic charge-$2e$ order, where the dashed line denotes the transition temperature $T_{\text{dw}}=4.05$.
}
\label{J1d2_trans}
\end{figure}

\begin{figure}[h]
\includegraphics[trim=0 0 0 0, clip,width=0.8 \columnwidth]{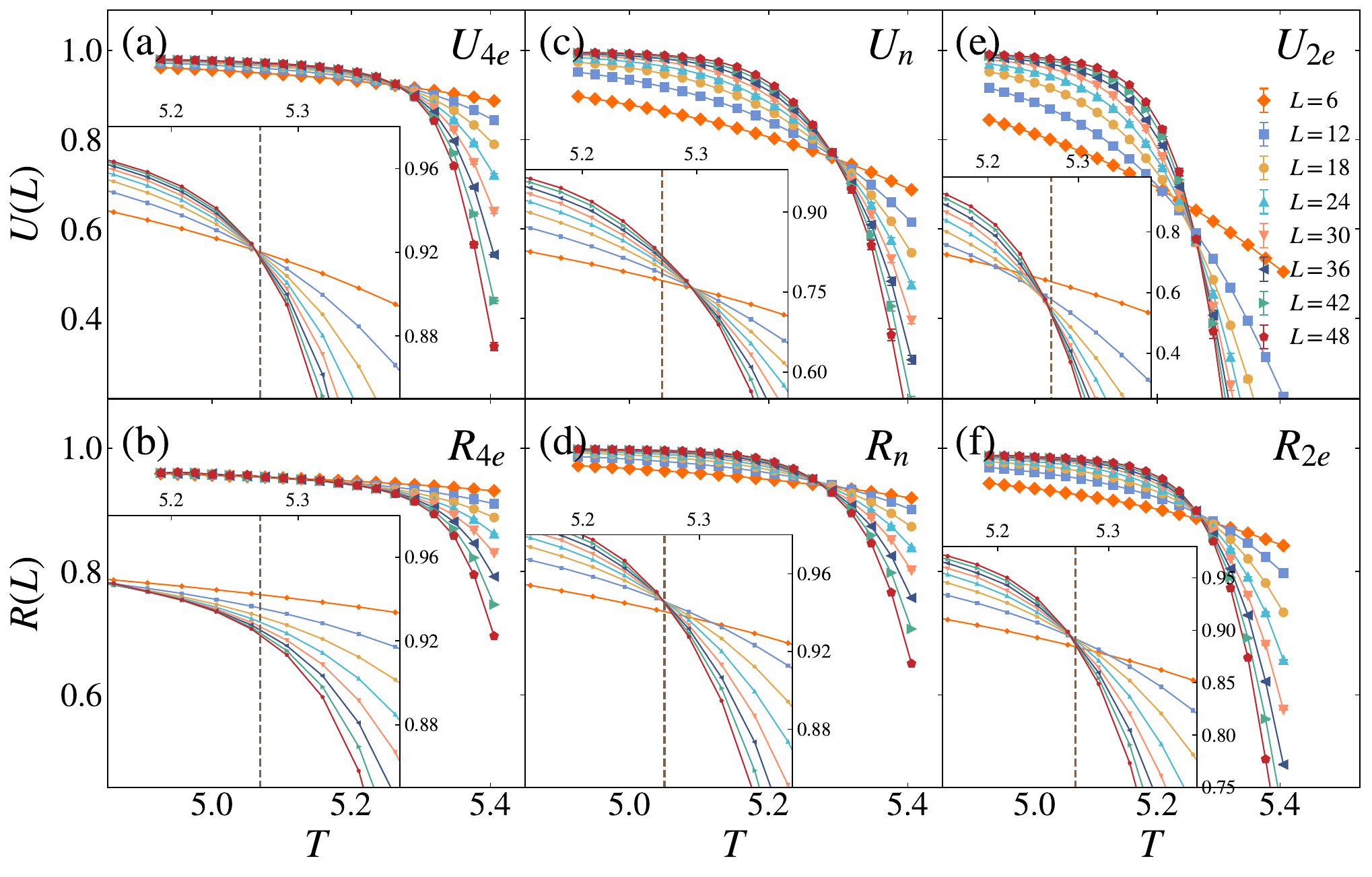}
\caption{
The plot of Binder cumulants $U(L)$ and the RG-invariant ratio $R(L)$ for the different orders at $J_3=1.8$. The insets depict a close-up view of the transition points. Panels (a)-(b), (c)-(d), and (e)-(f) correspond to the charge-$4e$, nematic, and isotropic charge-$2e$ orders, respectively. In all panels, the dashed line marks the common transition temperature $T=5.27$.
}
\label{J1d8_trans}
\end{figure}

In the main text, we present the corresponding figures for $J_3=0.4$. Here, we further provide the results for $J_3=1.2$ and $J_3=1.8$ in Figs.~\ref{J1d2_trans} and \ref{J1d8_trans}, respectively.
As shown in Table~\ref{tab1}, the transition temperatures obtained from the spin stiffness $\rho$ are in close agreement with those from $U_{4e}$ for all $J_3$, and also consistent with the results obtained from $U_n$ within numerical error for $J_3>J_3^*$, where $0.8<J_3^*<1.0$. The transition points obtained from $U_{2e}$ also coincide with $R_n$ within numerical error for all $J_3$. As illustrated in Figs.~\ref{J1d2_trans} and \ref{J1d8_trans}, the results at $J_3=1.2$ clearly show two distinct transitions, whereas those at $J_3=1.8$ are consistent with a single direct transition.


\subsection{Real-space evidence for type-(c) domain-wall proliferation}
\begin{figure}[t]
\includegraphics[trim=0 0 0 0, clip,width=1.0 \columnwidth]{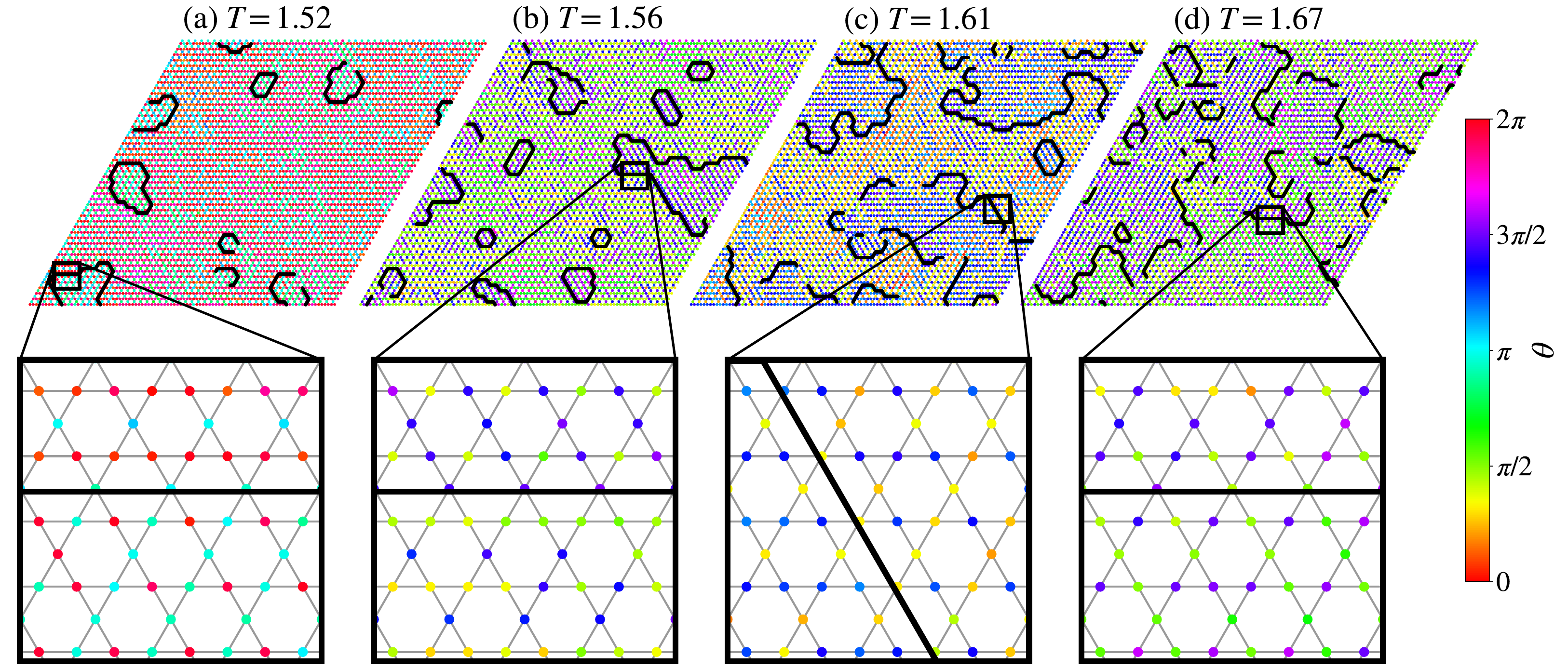}
\caption{
Real-space snapshots of domain-wall configurations at $J_3=0.4$ for an $L=48$ system across the transition temperature $T_{\mathrm{dw}}=1.59$: (a) $T=1.52$, (b) $T=1.56$, (c) $T=1.61$, and (d) $T=1.67$. The top row shows the full configurations, and the bottom row shows enlarged views of the boxed regions. The site colors encode the local phase $\theta$, and the black solid lines mark the identified domain walls. At low temperature the domain walls are sparse and short, while above $T_{\mathrm{dw}}$ they proliferate strongly and form extended connected structures. The enlarged views show that the dominant local change across the walls is of type-(c). Because $\theta$ is a continuous $U(1)$ variable, the identification of domain boundaries in thermal configurations is not unique; the black lines therefore indicate the dominant domain structure and are intended mainly as a visual guide.
}
\label{domain}
\end{figure}
To provide direct real-space evidence for the domain-wall-driven mechanism discussed in the main text, Fig.~\ref{domain} shows representative thermal snapshots of the site phases at $J_3=0.4$ on an $L=48$ kagome lattice for temperatures across the transition at $T_{\mathrm{dw}}=1.59$. The site colors encode the local phase $\theta$, while the black solid lines indicate the identified domain walls. At low temperature, Fig.~\ref{domain}(a) with $T=1.52$, the domain walls are sparse and short, consistent with the ordered phase. As the temperature approaches $T_{\mathrm{dw}}$ from below, Fig.~\ref{domain}(b) with $T=1.56$, both the number and the typical length of the domain walls increase. Above the transition, Figs.~\ref{domain}(c) and \ref{domain}(d) with $T=1.61$ and $1.67$, the domain walls proliferate strongly and form extended connected structures across the system. This evolution is fully consistent with the picture that the phase-I to phase-II transition is driven by the proliferation of domain walls.

The enlarged views in the lower panels further show that the proliferating walls are of the expected type-(c). For example, in the enlarged region of Fig.~\ref{domain}(a), the phase pattern on the three kagome sublattices across a representative wall changes from approximately $(\theta_x,\theta_y,\theta_z)\sim(\mathrm{blue},\mathrm{red},\mathrm{red})$ on one side to $(\theta_x,\theta_y,\theta_z)\sim(\mathrm{blue},\mathrm{blue},\mathrm{red})$ on the other side, which is precisely the type-(c) structure discussed in the main text.

Since $\theta$ is a continuous variable, the domain boundaries are not always uniquely defined. The black lines, therefore, indicate only the dominant domain structure. In addition, small domains and local fluctuating defects can make the configurations on the two sides of a domain wall deviate from an ideal pattern. Nevertheless, the overall proliferation of type-(c) domain walls across $T_{\mathrm{dw}}$ is clear and provides direct microscopic support for the domain wall-driven transition from phase I to phase II.

\subsection{Critical-exponent analysis}
In the quasi-long-range order phase, the magnetization satisfies $\ln(m^2)\sim -\eta\ln(L)+\text{const}.$ 
where $m^2=\frac{1}{N^2}\langle|\sum_i\Delta_{2e}(\mathbf{r}_i)|^2\rangle$ for the charge-$2e$ case, and is defined analogously for the charge-$4e$ and nematic order parameters.

\begin{figure}[t]
\includegraphics[trim=60 40 60 80, clip,width=0.9 \columnwidth]{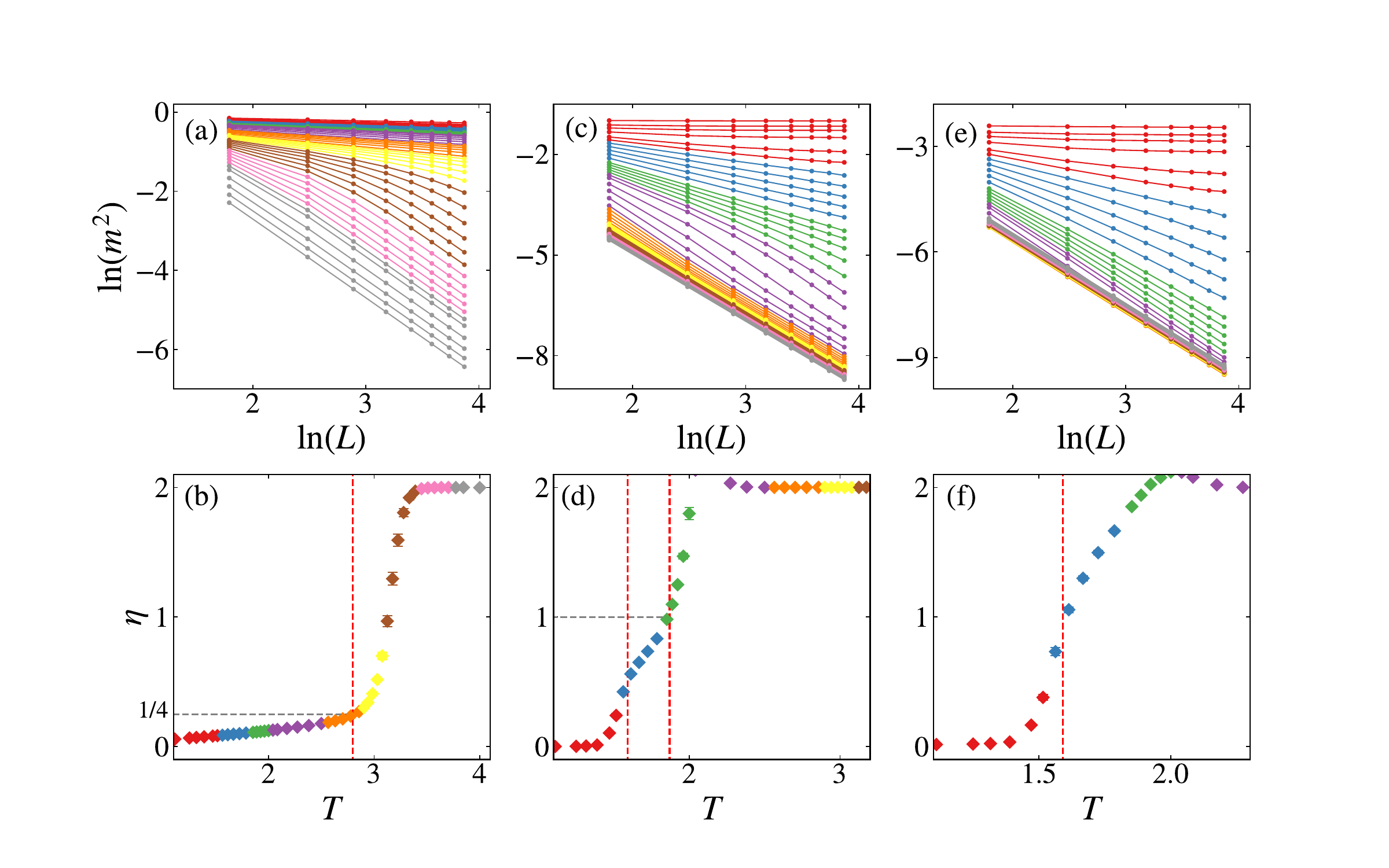}
\caption{The plot of $\ln(m^2)$ vs. $ \ln(L)$ and the critical exponents $\eta$ vs. $T$ for $J_3=0.4$, where $m^2$ is defined with (a-b) charge-$4e$, (c-d) nematic, and (e-f) charge-$2e$ order parameters. 
The colors in (a), (c), and (e) represent different temperatures as specified in (b), (d), and (f). 
The red dashed lines in (b), (d), and (f) denote the transition points $T=2.7971$, $(1.87,1.59)$, and $1.59$.}
\label{J0d4_eta}
\end{figure}

\begin{figure}[t]
\includegraphics[trim=60 40 60 80, clip,width=0.9 \columnwidth]{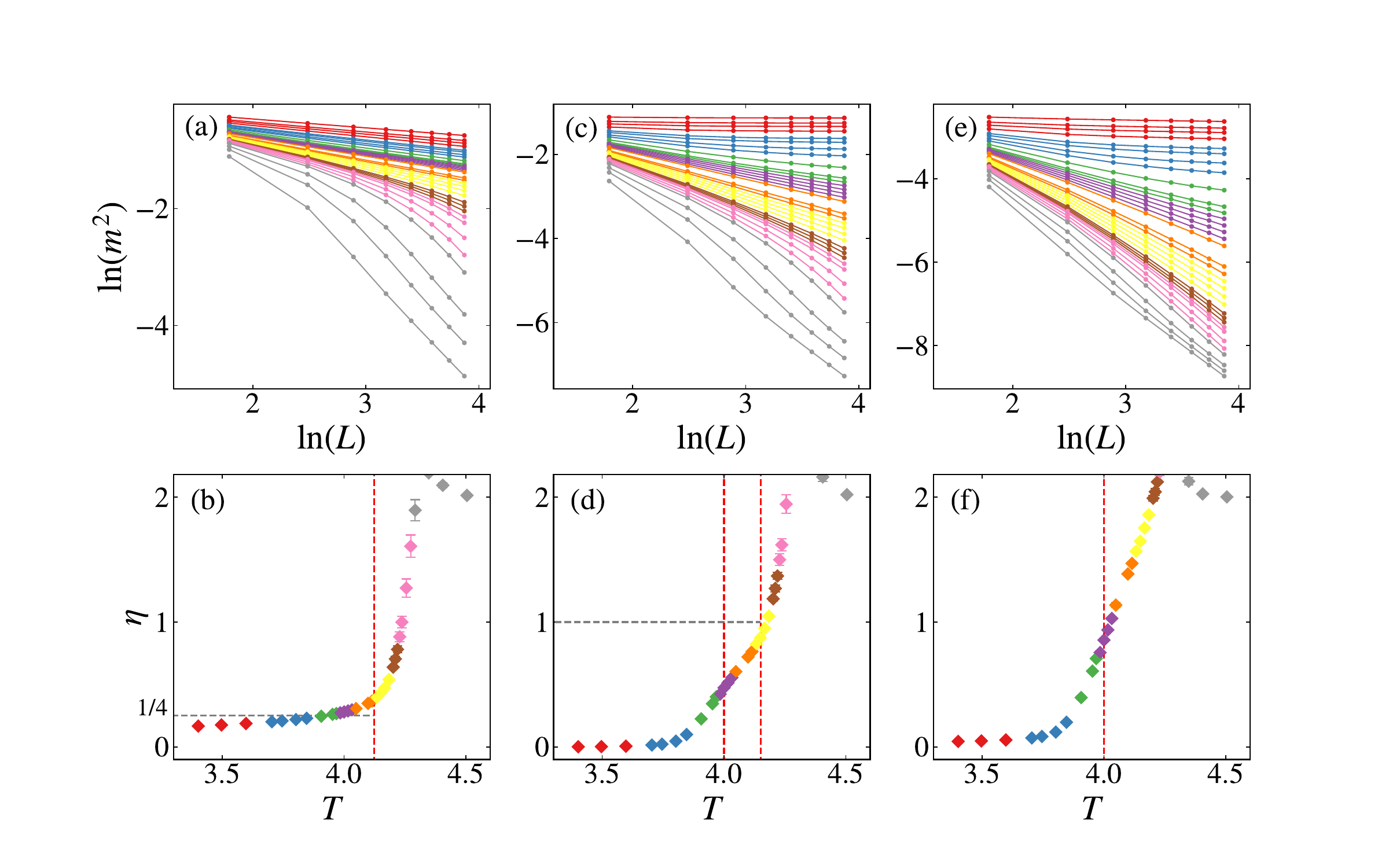}
\caption{
The plot of $\ln(m^2)$ vs. $ \ln(L)$ and the critical exponents $\eta$ vs. $T$ for $J_3=1.2$, where $m^2$ is defined with (a-b) charge-$4e$, (c-d) nematic, and (e-f) charge-$2e$ order parameters. 
The colors in (a), (c), and (e) represent different temperatures as specified in (b), (d), and (f). 
The red dashed lines in (b), (d), and (f) denote the transition points $T=4.123$, $(4.15,4.0)$, and $4.0$.}
\label{J1d2_eta}
\end{figure}

We apply linear fitting to the data with the four largest sizes $L=30\sim 48$ and obtain the exponents $\eta$ as shown in Fig.~\ref{J0d4_eta} and Fig.~\ref{J1d2_eta}.

For $J_3=0.4$, the behavior of the critical exponent $\eta$ for charge-4e order as a function of temperature is shown in Fig.~\ref{J0d4_eta} (b). As the temperature increases, $\eta$ increases slowly until it reaches the BKT transition temperature, where the value of $\eta_c=1/4$ is consistent with the $(\frac 1 2, 0)$ vortex proliferation (vortex for charge-$4e$ order parameters).

The critical exponent $\eta$ for the nematic order shows a two-step transition behavior, as shown in Fig.~\ref{J0d4_eta}(d). 
It is close to 0 at $T<1.59$ and converges to 2 at $T>1.87$. When $1.59<T<1.87$, the critical exponent $\eta$ increases slowly with the temperature and indicates the quasi-long-range order phase. 
At $T=1.87$, $\eta$ is very close to 1, which is consistent with nematic vortex proliferation.
As mentioned in the main text, the two elementary order parameters are $\Delta_\pm = |\Delta_{\pm}|e^{i\theta \pm \phi}$, and the nematic order parameter $Q\sim \Delta_{+}\Delta_{-}^*\sim e^{i 2\phi}$ is only related to the nematic field $\phi$.
In the nematic quasi-long-range order phase, the correlation function has a power law behavior $\langle \cos(\phi_i-\phi_j) \rangle \sim e^{-\langle (\phi_i-\phi_j)^2\rangle/2} \sim|\vec{R_i}-\vec{R_j}|^{-\tilde{\eta}(T)}$ with $\tilde{\eta}(T_c)=1/4$. Since $Q\sim e^{i 2\phi}$, the critical exponent $\eta_c$ should be four times larger here, which is equal to 1. The critical exponent $\eta$ for the isotropic charge-$2e$ order exhibits a transition near $T \simeq 1.59$, as shown in Fig.~\ref{J0d4_eta}(f).
 
For $J_3>J_3^*$, both nematic and charge-$4e$ quasi–long-range orders vanish simultaneously in a direct transition driven by $(\tfrac{1}{2},\tfrac{1}{2})$ vortex proliferation. Results for $J_3=1.2$ are shown in Fig.~\ref{J1d2_eta}. The critical exponents $\eta$ for the charge-$4e$ and nematic order parameters follow trends similar to those for $J_3<J_3^*$, but deviate from the expected values ($\eta=1/4$ for charge-$4e$ and $\eta=1$ for nematic). These deviations are natural consequences of $(\tfrac{1}{2},\tfrac{1}{2})$ vortex proliferation.

\begin{figure}[t]
\includegraphics[trim=0 0 0 0, clip,width=1.0 \columnwidth]{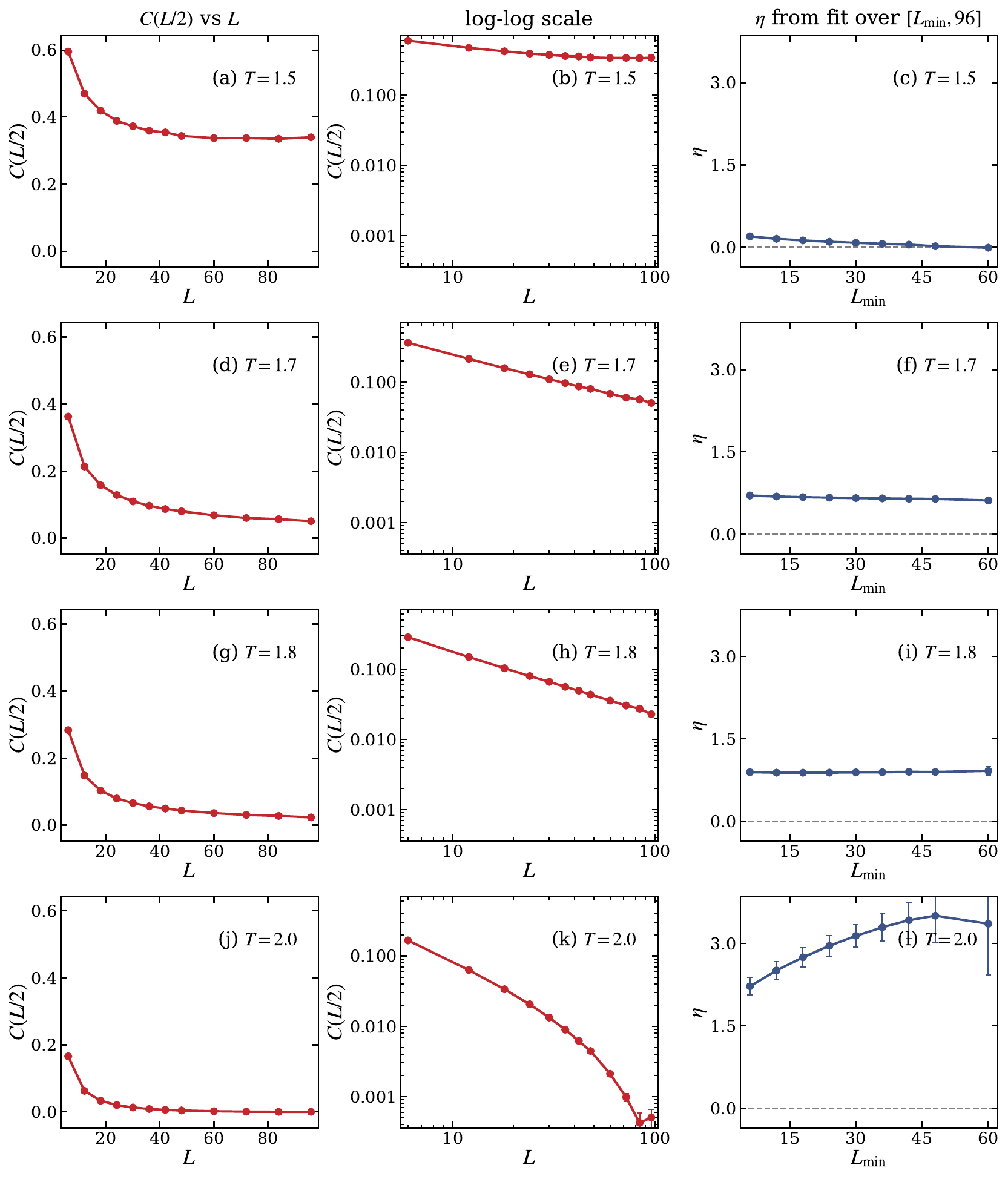}
\caption{
System-size dependence of the nematic correlation function at $J_3=0.4$ for four representative temperatures. The first column shows $C(L/2)$ versus $L$, the second column shows the same data on log-log scales, and the third column shows the exponent $\eta$ extracted from power-law fits $C(L/2)\sim L^{-\eta}$ over the interval $[L_{\min},96]$ as a function of $L_{\min}$. The four rows correspond to (a)-(c) $T=1.5$, (d)-(f) $T=1.7$, (g)-(i) $T=1.8$, and (j)-(l) $T=2.0$. Here $T=1.5$ is in phase I, $T=1.7$ and $1.8$ are inside the intermediate QLR nematic phase II, and $T=2.0$ is in the nematic-disordered phase III. 
}
\label{corrNemt}
\end{figure}

To further examine whether phase II is a genuine thermodynamic phase, we study the nematic correlation function at the largest distance, $C(L/2)$, for system sizes up to $L=96$ at $J_3=0.4$. The corresponding results are shown in Fig.~\ref{corrNemt} for four temperatures: $T=1.5$, $1.7$, $1.8$, and $2.0$. Here, $T=1.5$ lies in phase I, $T=1.7$ and $1.8$ lie inside the intermediate QLR nematic phase II, and $T=2.0$ lies in the nematic-disordered phase III.

The behavior of $C(L/2)$ clearly distinguishes these regimes. At $T=1.5$, $C(L/2)$ tends toward a finite value as $L$ increases, consistent with long-range nematic order in phase I. By contrast, at $T=1.7$ and $1.8$, $C(L/2)$ decreases with system size but remains well described by a power-law form over the accessible size range, as seen from the linear behavior in the log-log plots. To quantify this, we fit $C(L/2)\sim L^{-\eta}$ using data in the interval $[L_{\min},96]$ and examine the extracted exponent $\eta$ as a function of $L_{\min}$. In both cases, the fitted $\eta$ remains stable as $L_{\min}$ is varied, providing direct support for quasi-long-range nematic order in phase II. In contrast, at $T=2.0$, $C(L/2)$ decays much more rapidly with increasing $L$, and the effective exponent extracted from a power-law fit becomes strongly $L_{\min}$ dependent and grows to large values, indicating that a simple power-law description no longer applies. This behavior is consistent with the disordered phase.

Taken together, these results support the interpretation that the intermediate phase II is not a finite-size crossover region caused by the limited system sizes but instead is a distinct phase with quasi-long-range nematic order. 

\let\oldaddcontentsline\addcontentsline
\renewcommand{\addcontentsline}[3]{}
\bibliography{ref.bib}
\let\addcontentsline\oldaddcontentsline
\onecolumngrid